\documentclass[a4paper,twocolumn,numberedappendix,linenumbers]{emulateapj}
\usepackage[breaklinks,colorlinks=true,linkcolor=blue,anchorcolor=blue,citecolor=blue,urlcolor=blue,draft=false]{hyperref}
\usepackage{color}
\usepackage{amsmath} % align equations
\usepackage{amssymb}
\usepackage{mathrsfs}
\usepackage{placeins}
\usepackage{times}
\usepackage{xcolor}
\usepackage{xspace}
\usepackage{datetime}

\newcommand{\RN}[1]{%
  \textup{\uppercase\expandafter{\romannumeral#1}}%
}
\newcommand{\bp}{\mbox{\boldmath $p$}}
\newcommand{\bt}{\mbox{\boldmath $t$}}
\newcommand{\percent}{\ensuremath{\%}}

\usepackage{graphicx}
\usepackage[normalem]{ulem}

\usepackage{natbib}
\shorttitle{Likelihood-free Forward Modeling for Cluster Weak Lensing and Cosmology}
\shortauthors{Tam, Umetsu, \& Amara}

\begin{document}

\title{Likelihood-free Forward Modeling for Cluster Weak Lensing and Cosmology}

%\correspondingauthor{}
\email{sitam@asiaa.sinica.edu.tw}
\email{keiichi@asiaa.sinica.edu.tw}
\email{adam.amara@port.ac.uk}

%\author{Sut-Ieng Tam}
%\affiliation{Academia Sinica Institute of Astronomy and Astrophysics (ASIAA), No.~1, Sec.~4, Roosevelt Road, Taipei 10617, Taiwan}

%\author{Keiichi Umetsu}
%\affiliation{Academia Sinica Institute of Astronomy and Astrophysics (ASIAA), No.~1, Sec.~4, Roosevelt Road, Taipei 10617, Taiwan}

%\author{Adam Amara}
%\affiliation{Institute of Cosmology \& Gravitation, University of Portsmouth, Dennis Sciama Building, Burnaby Road, Portsmouth PO1 3FX, UK}

%%% arXiv@@
\author{Sut-Ieng Tam\altaffilmark{1}}
\author{Keiichi Umetsu\altaffilmark{1}}
\author{Adam Amara\altaffilmark{2}}
\altaffiltext{1}{Academia Sinica Institute of Astronomy and Astrophysics (ASIAA), No.~1, Sec.~4, Roosevelt Road, Taipei 10617, Taiwan}
\altaffiltext{2}{Institute of Cosmology \& Gravitation, University of Portsmouth, Dennis Sciama Building, Burnaby Road, Portsmouth PO1 3FX, UK}

\begin{abstract}
Likelihood-free inference provides a rigorous approach to preform Bayesian analysis using forward simulations only. The main advantage of likelihood-free methods is its ability to account for complex physical processes and observational effects in forward simulations. Here we explore the potential of likelihood-free forward modeling for Bayesian cosmological inference using the redshift evolution of the cluster abundance combined with weak-lensing mass calibration. We use two complementary likelihood-free methods, namely Approximate Bayesian Computation (ABC) and Density-Estimation Likelihood-Free Inference (DELFI), to develop an analysis procedure for inference of the cosmological parameters $(\Omega_\mathrm{m},\sigma_8)$ and the mass scale of the survey sample. Adopting an \textit{eROSITA}-like selection function and a $10\percent$ scatter in the observable--mass relation in a flat $\Lambda$CDM cosmology with $\Omega_\mathrm{m}=0.286$ and $\sigma_8=0.82$, we create a synthetic catalog of observable-selected NFW clusters in a survey area of $50$~deg$^2$. The stacked tangential shear profile and the number counts in redshift bins are used as summary statistics for both methods. By performing a series of forward simulations, we obtain convergent solutions for the posterior distribution from both methods. We find that ABC recovers broader posteriors than DELFI, especially for the $\Omega_\mathrm{m}$ parameter. For a weak-lensing survey with a source density of $n_\mathrm{g}=20$~arcmin$^{-2}$, we obtain posterior constraints on $S_8=\sigma_8(\Omega_\mathrm{m}/0.3)^{0.3}$ of $0.836\pm 0.032$ and $0.810\pm 0.019$ from ABC and DELFI, respectively. The analysis framework developed in this study will be particularly powerful for cosmological inference with ongoing cluster cosmology programs, such as the \textit{XMM}-XXL survey and the \textit{eROSITA} all-sky survey, in combination with wide-field weak-lensing surveys. 
\end{abstract}


\section{Introduction} \label{sec:intro}

As the largest bound objects formed in the universe, galaxy clusters play a fundamental role in testing models of background cosmology and structure formation. In the standard picture of hierarchical structure formation, the abundance of cluster halos as a function of mass and redshift is sensitive to the amplitude and growth rate of density fluctuations and the cosmic volume--redshift relation \citep[e.g.,][]{Haiman2001}. Cluster number counts measured over a wide range in mass and redshift can thus provide powerful cosmological constraints especially on the matter density parameter $\Omega_\mathrm{m}$ and the amplitude of linear density fluctuations $\sigma_{8}$ (defined in detail at the end of this section) \citep[e.g.,][]{2015MNRAS.446.2205M}. In this context, recent and ongoing cluster surveys covering a significant fraction of the sky allow us to place stringent constraints on the cosmological parameters \citep[e.g.][]{2016ApJ...832...95D,2017MNRAS.471.1370S,2018A&A...620A..10P,2019ApJ...878...55B,2021PhRvD.103d3522C,2021PhRvL.126n1301T,Chiu2021}. 

Cosmological parameters in the standard $\Lambda$ cold dark matter ($\Lambda$CDM) model derived from low-redshift cosmological probes, such as galaxy clusters and cosmic shear, are often in tension with those from observations of cosmic microwave background (CMB) anisotropies \citep{Planck2018VI}. Such apparent discrepancies in terms of $\Omega_\mathrm{m}$ and $\sigma_8$ are often referred to as the ``$S_8$ tension'' \citep[e.g.,][]{Hildebrandt2017}, where $S_8=\sigma_8 (\Omega_\mathrm{m}/0.3)^\alpha$ with $\alpha$ being a constant that depends on the degree of parameter degeneracy (typically, $\alpha=0.3$--$0.5$; see Section~\ref{sec:results}). 
In general, there are various challenging issues associated with cosmological tests using low-redshift probes, especially galaxy clusters, which involve complex measurement processes and modeling in the highly nonlinear regime of structure formation coupled with baryonic physics \citep{Pratt2019}. To obtain robust cosmological constraints from clusters in the present era of precision cosmology, one needs to conduct accurate statistical inference accounting for various observational and instrumental effects in modeling processes.

Accurate calibration of cluster mass measurements is another critical ingredient of cluster cosmology \citep{Pratt2019}. In cluster surveys, different observational techniques are employed to define an observable-selected cluster sample using a low-scatter proxy (e.g., X-ray luminosity and temperature, integrated Compton $y$ parameter, and optical richness) that correlates with the underlying cluster mass.  
With the assumption of hydrostatic equilibrium or virial theorem, these mass proxies can provide cluster mass estimates, which however are expected to be biased by the presence of merging substructures, non-gravitational processes, or instrumentation effects \citep{Nagai2007,2014ApJ...794..136D}. Consequently, cosmological cluster studies often require an external mass calibration of the survey sample using direct mass measurements \citep{2016A&A...594A..24P,2018A&A...620A..10P}.

Weak gravitational lensing offers a direct probe of the total mass distribution around galaxy clusters projected along the line of sight, irrespective of their dynamical state \citep[e.g.][]{vonderLinden2014,Umetsu2014,2015MNRAS.449..685H,Okabe+Smith2016,Medezinski2018,2019MNRAS.483.2871D,2020MNRAS.497.4684H,2020MNRAS.496.4032T,Chiu2021}.  Cluster weak lensing thus allows us to obtain an unbiased mass calibration of galaxy clusters for accurate cosmology, if systematic effects, such as shear calibration bias, photometric redshift bias, and mass modeling bias, are under control \citep{Pratt2019}. 

In cluster cosmology, a Bayesian statistical approach is often used to derive cosmological parameter constraints from observational data, because the Bayesian framework enables probabilistic incorporation of prior knowledge about uncertain physical processes. This framework assumes that the likelihood of data given a set of model parameters is known. In practice, a Gaussian likelihood is often assumed. However, non-Gaussian contributions could dominate the errors owing to complex and nonlinear measurement processes. Moreover, statistical fluctuations of cluster properties at fixed halo mass (e.g., cluster lensing signals; \citealt{Gruen2015}) are likely non-Gaussian due to their nonlinear nature. As a result, Gaussian distributions are likely an insufficient representation for modeling cluster observations, so that the likelihood is essentially intractable. As a possible solution to overcome these difficulties, a simulation-based likelihood-free approach is receiving increasing attention.

In particular, \citet{2015A&C....13....1I} explored the utility of likelihood-free inference for cosmological analysis based on number counts of galaxy clusters selected from a Sunyaev--Zel'dovich (SZ) effect survey.
They used \textsc{numcosmo} \citep{numcosmo} to create a synthetic catalog of SZ-selected clusters from forward simulations, taking into account the uncertainties from photometric-redshift measurements and lognormal scatter in the SZ detection significance. Using SZ cluster counts combined with the distribution of cluster redshift and SZ detection significance as observable features, they demonstrated the possibility of using likelihood-free techniques for cluster cosmology.

In this paper, we aim to develop a likelihood-free procedure for accurate cosmological parameter inference based on the redshift evolution of the cluster abundance in combination with weak-lensing mass calibration. Specifically, we will use two different likelihood-free algorithms, namely Approximate Bayesian Computation \citep[ABC;][]{10.2307/2240995} and Density-Estimation Likelihood-Free Inference \citep[DELFI;][]{2012arXiv1212.1479F,2016arXiv160506376P,2017arXiv171101861L,2018arXiv180507226P, 2018arXiv180509294L,2018MNRAS.477.2874A}. 
ABC methods sample the model parameter space and compare simulated and observed datasets using a distance metric. Accepting parameter samples for which this distance is smaller than a given threshold, ABC provides an approximate posterior distribution of the model parameters. By contrast, DELFI requires much fewer simulations than ABC. It trains a set of neural density estimators for a target posterior by using simulated data--parameter pairs. These likelihood-free approaches allow us to bypass the need for an evaluation of the likelihood by using synthetic data made through forward modeling. In this study, we will use two publicly available software packages, 
\textsc{abcpmc} \citep{abcpmc} and \textsc{pydelfi} 
\citep{2019MNRAS.488.4440A}, 
which implement the ABC and DELFI algorithms respectively. We note that, in contrast to this work, \citet{2015A&C....13....1I} used the SZ mass proxy and redshift as cluster observables, focusing on an ABC algorithm (\textsc{cosmoabc}).

This paper is organized as follows. The formalism of cluster--galaxy weak lensing and the modeling procedure of our forward simulations are described in Section~\ref{sec:methods}.  Section~\ref{sec:simulation} summarises the likelihood-free inference methods. In Section~\ref{sec:results}, we present the results of  likelihood-free cosmological inference and discuss the prospects and current limitations of using our forward simulator for cosmological cluster surveys. Finally, we present our conclusions in Section~\ref{sec:conclusion}.

Throughout this paper, we assume a spatially flat $\Lambda$CDM cosmology with $\Omega_\mathrm{m}=0.286$, $\Omega_\Lambda=0.714$, a Hubble constant of $H_0=100~h$~km~s$^{-1}$~Mpc$^{-1}$ with $h=0.7$, and $\sigma_8=0.82$ \citep{Hinshaw2013}, where $\sigma_8$ is the rms amplitude of linear density fluctuations in a sphere of comoving radius $8 h^{-1}$~Mpc at $z=0$. We denote the critical density of the universe at a particular redshift $z$ as $\rho_\mathrm{c}(z)=3H^2(z)/(8\pi G)$, with $H(z)$ the redshift-dependent Hubble function. We adopt the standard notation $M_\Delta$ to denote the mass enclosed within a sphere of radius $r_\Delta$ within which the mean overdensity equals $\Delta \times \rho_\mathrm{c}(z)$. We denote three-dimensional cluster radii as $r$ and reserve the symbol $R$ for projected cluster-centric distances. We use "$\log$" to denote the base-10 logarithm and "$\ln$" to denote the natural logarithm. The fractional scatter in natural logarithm is quoted as a percent. All quoted errors are $1\sigma$ confidence levels unless otherwise stated.

\section{Modeling Procedure}
\label{sec:methods}

\subsection{Basics of Cluster Weak Lensing}

Weak gravitational lensing causes small but coherent distortions in the images of source galaxies lying behind overdensities such as galaxy clusters \citep[for a didactic review of cluster weak lensing, see][]{Umetsu2020rev}. The lensing convergence $\kappa$ is responsible for isotropic magnification and proportional to the surface mass density $\Sigma$ projected along the line of sight, 
\begin{equation}
\kappa = \Sigma/\Sigma_\mathrm{cr}
\label{eq:kappa}
\end{equation}
with $\Sigma_\mathrm{cr}$ the critical surface mass density for gravitational lensing as a function of lens redshift $z_l$ and source redshift $z_s$, defined as
\begin{equation}
\Sigma_\mathrm{cr}(z_l, z_s) =\frac{c^2 D_s(z_s)} {4\pi G D_l(z_l) D_{ls}(z_l,z_s)},
\end{equation}
where
$D_l(z_l)$, $D_s(z_s)$, $D_{ls}(z_l,z_s)$ are the angular diameter distances from the observer to the lens, from the observer to the source, and from the lens to the source, respectively. For an unlensed source with $z_s\le z_l$, $\Sigma_\mathrm{cr}^{-1}(z_l,z_s)=0$.

The shape distortion due to lensing is described by the complex gravitational shear,
\begin{equation}
\gamma  = \gamma_1+i\gamma_2.
\label{eq:gamma_ini}
\end{equation}
The observable quantity for weak shear lensing is the reduced shear,
\begin{equation}
    g :=g_1+ig_2 = \frac{\gamma}{1-\kappa},
\end{equation}
which can be directly estimated from the image ellipticities of background galaxies.

The shear ($\gamma_1,\gamma_2$) can be decomposed into the tangential component $\gamma_{+}$ and the 45$^{\circ}$-rotated cross component $\gamma_{\times}$ defined with respect to the cluster center. The azimuthally averaged tangential shear $\gamma_+(R)$ as a function of cluster radius $R$ is proportional to the excess surface mass density $\Delta\Sigma(R)$, defined as
\begin{equation}
    \Delta\Sigma(R)\equiv\Sigma(<R)-\Sigma(R)=\Sigma_\mathrm{cr}\,
    \gamma_{+}(R),
    \label{eq:gmmat}
\end{equation}
where $\Sigma(R)$ represents the azimuthally averaged surface mass density at cluster radius $R$ and $\Sigma(<R)$ is its mean interior to the radius $R$.

The reduced tangential shear signal $g_+(R)$ as a function of cluster radius is related to $\Sigma(R)$ and $\Delta\Sigma(R)$ as
\begin{equation}
\label{eq:gt}
 g_+(R) = \frac{\Sigma_\mathrm{cr}^{-1}\Delta\Sigma(R)}{1-\Sigma_\mathrm{cr}^{-1}\Sigma(R)}.
\end{equation}
The azimuthally averaged cross-shear component, $g_\times(\theta)$, is expected to vanish if the signal is caused by weak lensing. 
 
\subsection{Cluster Abundance and Stacked Weak-lensing Signal}
\label{subsec:obs}

\begin{figure}[htbp]
 \centering
 \includegraphics[width=0.47\textwidth,clip]{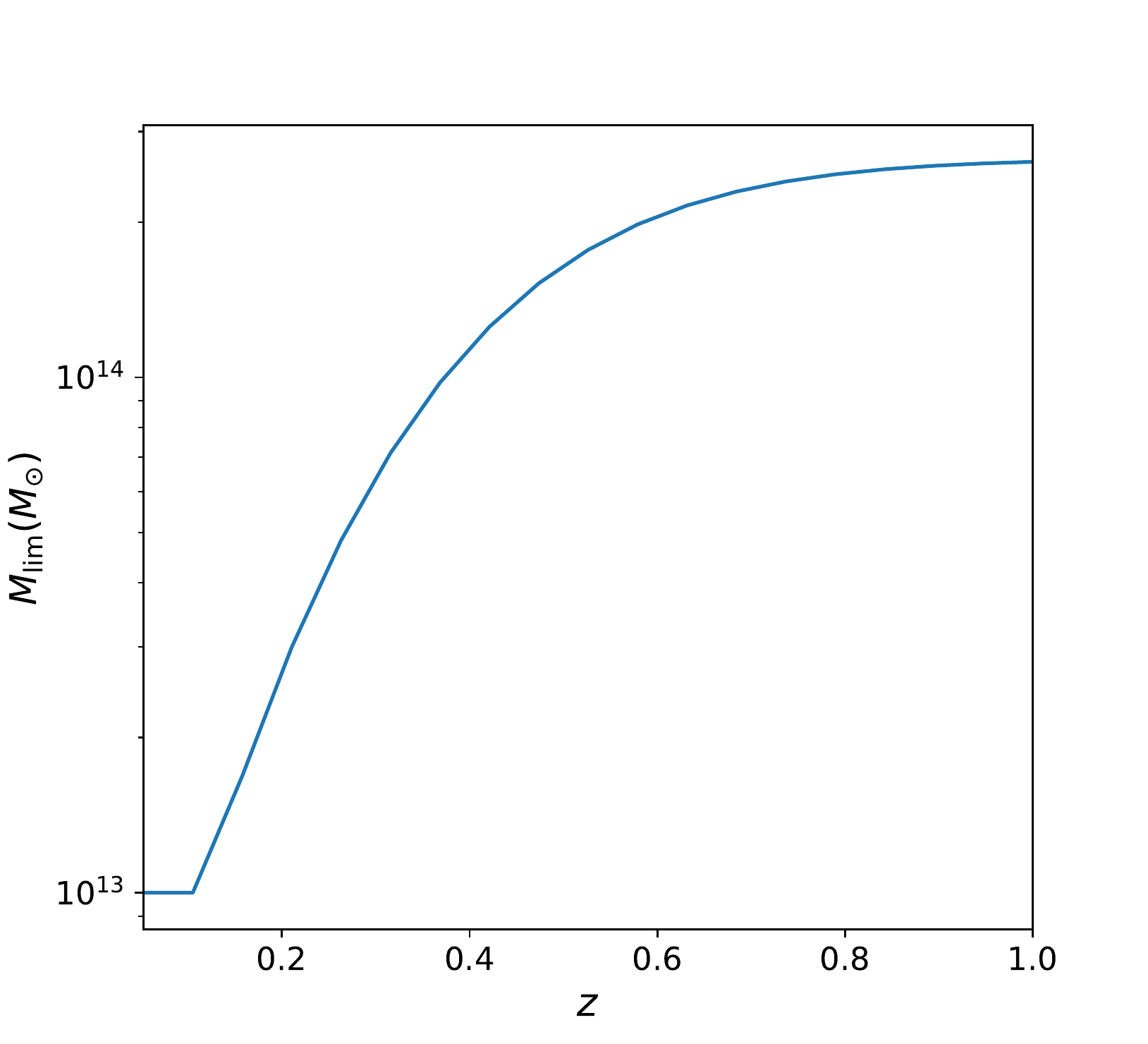}
 \caption{Minimum mass threshold $M_\mathrm{lim}(z)$ as a function of redshift assumed in this study. We only consider clusters at $z<1$. This $M_\mathrm{lim}(z)$ function mimics the \textit{eROSITA} selection function with a detection threshold of 50 photon counts.}
    \label{fig:selection}
\end{figure}

For a given cosmology and a given survey selection function, the abundance of galaxy clusters detected by the survey can be predicted. The redshift distribution of galaxy clusters detected by a survey is expressed as
\begin{equation}
\begin{aligned}
    \frac{dN_\mathrm{cl}}{dz}(z) &= 4\pi f_\mathrm{sky}
    \frac{cr^2(z)}{H(z)}
    \int\! dM\,\frac{dn(M,z)}{dM}\\
    &\times \int\!d\ln{M'}S(\ln{M'},z)\,P(\ln{M'}|\ln{M}), 
    \label{eq:dndz}
    \end{aligned}
\end{equation}
where $f_\mathrm{sky}=\Omega_\mathrm{s}/(4\pi)$ is the sky coverage fraction with $\Omega_\mathrm{s}$ the solid area of the survey, $dn(M,z)/dM$ is the comoving mass function of halos, $S(\ln{M'},z)$ is the survey selection function of the "observable" mass $\ln{M'}$, $P(\ln{M'}|\ln{M})$ is the conditional probability distribution function of $\ln{M'}$ for a given true logarithmic mass $\ln{M}$, and $cr^2(z)/H(z)$ is the comoving volume per unit redshift interval and per unit steradian. Here  $r=f_K(\chi)$ is the comoving angular diameter distance, with $f_K(\chi)=\chi$ for zero spatial curvature, $K=0$. The total number of clusters detected by the survey is $N_\mathrm{cl}=\int\!dz\, dN_\mathrm{cl}(z)/dz$.

In this study, we adopt the halo mass function given by \cite{2016MNRAS.456.2486D} with the halo mass definition of $M_{500}$. We assume a selection function of the form
\begin{equation}
    S(\ln{M'})= {\cal H}\left[\ln{M'}-\ln{M_\mathrm{min}(z)}\right]
\end{equation}
where $M_\mathrm{min}(z)$ is the minimum mass threshold as a function of redshift and ${\cal H}(x)$ is the Heaviside step function defined such that ${\cal H}(x)=1$ for $x\ge 0$ and ${\cal H}(x)=0$ otherwise. The probability function $P(\ln{M'}|\ln{M})$ is assumed to be a Gaussian distribution with $\log{M'}=\log{M} \pm \sigma_\mathrm{int}$ with $\sigma_\mathrm{int}$ the intrinsic dispersion.
%\footnote{It is straightforward to generalize the observable--mass scaling relation, for example, to include the slope and intercept parameters as $\log{M'}=\alpha + \beta\log{M} \pm \sigma_\mathrm{int}$ \citep[e.g.,][]{Umetsu2020hsc}.}
We adopt a $10\percent$ intrinsic scatter in the observable--mass relation of $\sigma_\mathrm{int}=0.1/\ln{10}$. In this way, we take into account the effect of Eddington bias as well as statistical fluctuations in $\log{M'}$ on the selected sample of galaxy clusters. In Appendix~\ref{sec:mscale}, we present a general procedure for modeling the observable-mass relation.

In real observations, galaxy clusters are selected by their mass proxy from optical, X-ray, or SZ-effect observations. Here we assume an X-ray cluster survey over a total sky area of $\Omega_\mathrm{s}=50$~deg$^2$ (e.g., the XXL survey with the {\it XMM-Newton} X-ray satellite; see \citealt{XXL}). We adopt an \textit{eROSITA}-like selection function with the minimum mass threshold $M_\mathrm{lim}(z)$ parameterized as
\begin{equation}
\label{eq:mmin}
    \log
    \left[\frac{M_\mathrm{500,min}(z)}{M_\odot}\right]=
    \max\left\{13,
    A\left[1+\mathrm{erf}\left(\frac{z-B}{C}\right)\right]\right\}
\end{equation}
for $z<1$, with $A=7.212$, $B=-0.432$, and $C=0.602$. In this study, we set $M_\mathrm{500,min}(z)\to \infty$ at $z\ge 1$. Here $A$ sets the normalization of the $M_\mathrm{lim}(z)$ function, while $B$ and $C$ describe its redshift evolution. Figure~\ref{fig:selection} shows the cluster selection function in terms of $M_\mathrm{500,min}(z)$ adopted in this study. The fitting function given by Equation~(\ref{eq:mmin}) approximates well the \textit{eROSITA} selection function for a detection threshold of 50 photon counts \citep[][see their Figure~2]{2012MNRAS.422...44P}. We note that the selection function defined with Equation~(\ref{eq:mmin}) ensures that halos with $M^\prime_{500} <10^{13}M_\odot$ are not detected.

We model the mass distribution of individual cluster halos with a spherical Navarro--Frenk--White  (NFW) profile motivated by cosmological simulations of collisionless CDM \citep[][]{NFW1, NFW2}. This assumption is supported by observational and theoretical studies, which found that the stacked $\Delta\Sigma(R)$ profile around galaxy clusters can be well described by a projected NFW profile \citep[e.g.,][]{Oguri+Hamana2011,Okabe+Smith2016,Umetsu2016,Umetsu+Diemer2017,Sereno2017}.\footnote{The contribution from the 2-halo term to the excess surface density $\Delta\Sigma$ becomes significant at about several virial radii (see Figure~2 of \citealt{Oguri+Hamana2011}). In this study, we neglect the density steepening associated with the splashback radius.}

The NFW density profile is given by
\begin{equation}
\rho(r)=\frac{\rho_\mathrm{s}}{(r/r_\mathrm{s})(1+(r/r_\mathrm{s}))^2},
\end{equation}
where $\rho_\mathrm{s}$ is the characteristic density and $r_\mathrm{s}$ is the scale radius at which the logarithmic density slope equals $-2$.  We parametrize the NFW model with the halo mass $M_\Delta$ and the concentration parameter $c_{\Delta}\equiv r_{\Delta}/r_\mathrm{s}$ defined at $\Delta=500$. 
For a given cosmology, we assign a concentration to each cluster in our sample
using the mean concentration--mass ($c$--$M$) relation $c_{500}(M_{500},z|\Omega_\mathrm{m},\sigma_8)$ of \cite{Diemer19}. It should be noted that for the sake of simplicity, our modeling procedure neglects the effect of intrinsic scatter in the $c$--$M$ relation.\footnote{The concentration scatter inferred from lensing for X-ray-selected cluster samples is $\sim 20\percent$ \citep[see][]{Umetsu2020rev}, which is much lower than found for CDM halos in $N$-body simulations ($\sim 35\percent$ for the full population of halos including both relaxed and unrelaxed systems; see \citealt[][]{Bhatt2013,DK2015}.).} We will discuss in Section~\ref{subsec:assumptions} the implications of the assumptions made in the present study.

The stacked weak-lensing signal averaged over the sample of all detected clusters is written as 
\begin{equation}
\label{eq:stackedgt}
\begin{aligned}
\langle g_+\rangle(R_i) &= \frac{1}{N_\mathrm{cl}}\int\!dz\,\frac{cr^2(z)}{H(z)}
\int\!dM\,\frac{dn(M,z)}{dM}\\
\times& \int\!d\ln{M'}\,S(\ln{M'},z) \,P(\ln{M'}|\ln{M})\\
\times& g_+(R_i|M,z),
\end{aligned}
\end{equation}
where $g_+(R_i|M,z)$ is the expected reduced tangential shear signal in the $i$th radial bin ($i=1,2,\dots,N_\mathrm{bin}$)
for a cluster with halo mass $M$ and redshift $z$ (see Equation~(\ref{eq:gt})). 
To simplify the procedure and facilitate the interpretation of results, we assume that all source galaxies lie at a redshift of $z_s=1$, the typical mean redshift of spatially resolved background galaxies from deep ground-based imaging observations \citep[e.g.,][]{Umetsu2014}.
We note that Equation~(\ref{eq:stackedgt}) assumes the use of uniform weighting for lens--source pairs. It is straightforward to implement a redshift-dependent weighting for lensing \citep{Umetsu2014,Miyatake2019}. 

The dominant source of noise in weak shear lensing is the shape noise of background galaxy images. Assuming a shape dispersion of $\sigma_{g}=0.4/\sqrt{2}$ per galaxy per shear component, we add random-phase Gaussian noise with zero mean and dispersion $\sigma_{g,\mathrm{eff}}=\sigma_{g}/\sqrt{N_\mathrm{gal}}$ to the reduced tangential shear signal $g_+(R)$ for each cluster and each radial bin. Here $N_\mathrm{g}$ is the expected number of source galaxies in each radial bin $[R_i,R_{i+1}]$, $N_\mathrm{g}=\pi n_\mathrm{g}(R^2_{i+1}-R^2_{i})/D_l^2$, with $n_\mathrm{g}$ the mean surface number density of background galaxies. In addition, cosmic noise covariance arises from the projected large-scale structure uncorrelated with the clusters \citep{Schneider1998,2003MNRAS.339.1155H}. This noise is correlated between radial bins and becomes important at large cluster distances where the cluster lensing signal is small \citep{Miyatake2019}. We thus neglect the cosmic noise contribution in this study. In principle, it is straightforward to compute the cosmic noise covariance matrix $C^\mathrm{lss}$ for a given cosmology using the nonlinear matter power spectrum \citep[see][]{Oguri+Takada2011,Umetsu2020rev}. We also neglect the contribution from statistical fluctuations of the cluster lensing signal ($C^\mathrm{int}$) due to intrinsic variations associated with assembly bias and cluster asphericity \citep[see][]{Gruen2015, Umetsu2016, Miyatake2019, Umetsu2020rev}.

In this study, we consider two different weak-lensing sensitivities defined in terms of the background galaxy density parameter $n_\mathrm{g}$, namely $n_\mathrm{g}=20$~galaxies~arcmin$^{-2}$ and $n_\mathrm{g}=400$~galaxies~arcmin$^{-2}$. Our fiducial analysis uses $n_\mathrm{g}=20$~galaxies~arcmin$^{-2}$, which is close to the typical value of $n_\mathrm{g}$ for weak-lensing shape measurements with the 8.2~m Subaru telescope \citep[e.g.,][]{Miyatake2019,Umetsu2020hsc}.\footnote{Applying a background selection based on color and photometric-redshift information, the typical number density of background galaxies for cluster weak lensing is reduced to $n_\mathrm{g}=12$--$14$~galaxies~arcmin$^{-2}$ \citep[e.g.,][]{Umetsu2014,Medezinski2018}.}
The case with $n_\mathrm{g}=400$~galaxies~arcmin$^{-2}$ represents an idealized, essentially "noise-free" setup for comparison purposes. For the latter case, the uncertainties in the inferred parameters will be dominated by the sample variance of the selected sample, in contrast to the former case with noisy weak-lensing measurements. The comparison of the two different realizations will thus give us a crude idea of the respective error contributions to the parameter uncertainty.

Finally, we simulate reduced tangential shear profiles $\{g_{+}\}_{i=1}^{N_\mathrm{bin}}$ for all clusters in $N_\mathrm{bin}=10$ equally spaced logarithmic bins of comoving cluster radius $R$, ranging from $R_\mathrm{min}$ = $0.3\,h^{-1}$~Mpc to $R_\mathrm{max}$ = $3\,h^{-1}$~Mpc typically adopted in cluster weak-lensing studies with Subaru Hyper Suprime-Cam observations \citep[e.g.,][]{Umetsu2020hsc}. For our fiducial choice of the weak-lensing sensitivity with $n_\mathrm{g}=20$~galaxies~arcmin$^{-2}$, the contributions from both $C^\mathrm{lss}$ and $C^\mathrm{int}$ can be safely ignored within the chosen radial range \citep{Miyatake2019,Umetsu2020rev}.

\section{Likelihood-free Forward Modeling}
\label{sec:simulation}

\begin{figure}[htbp]
 \centering
 \includegraphics[width=0.48\textwidth,clip]{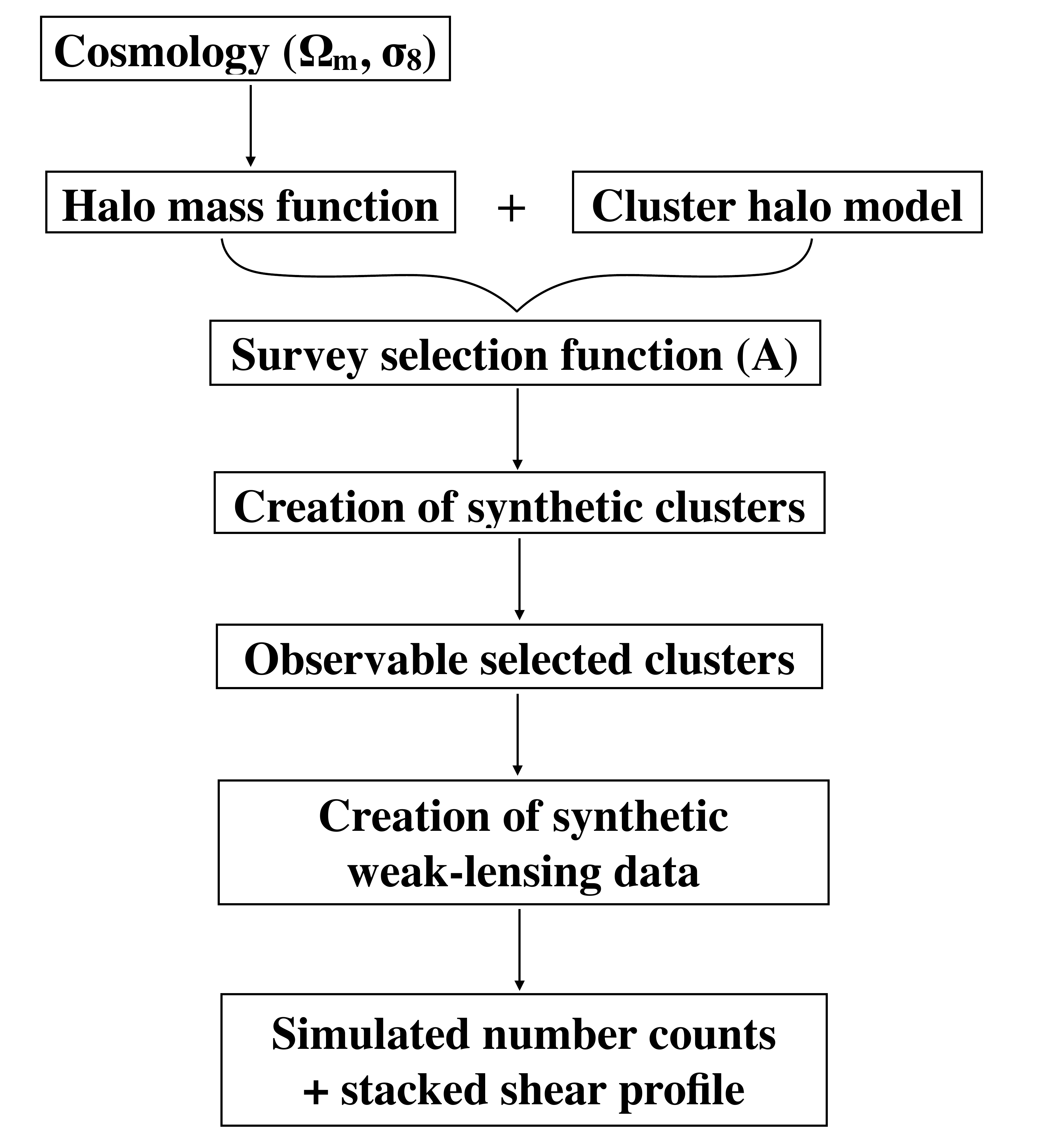}
 \caption{Schematic diagram illustrating the forward modelling procedure for our cosmological inference pipeline. The fiducial parameters for the cluster survey considered in this paper are $\Omega_\mathrm{m} = 0.286$, $\sigma_8=0.82$, and $A=7.212$.}
 \label{fig:flowchart}
\end{figure}

A schematic diagram of our forward-modeling procedure is shown in Figure~\ref{fig:flowchart}. First, our forward model $F(\Omega_\mathrm{m},\sigma_8,A)$ is specified for a given set of $(\Omega_\mathrm{m},\sigma_8)$ and for a given value of the $A$ parameter that sets the normalization of $M_\mathrm{lim}(z)$, corresponding to the mass scale of the survey sample (see Appendix~\ref{sec:mscale} for a general treatment of the mass calibration). Our fiducial model is $F(\Omega_\mathrm{m}=0.286, \sigma_8=0.82, A=7.212)$. In this work, we adopt the mass function by \citet{2016MNRAS.456.2486D} and an NFW halo description with the $c(M,z)$ relation by \citet{Diemer19}, both of which depend on the input cosmology. Next, we construct a synthetic catalog of halos with $M_{500}>10^{11}h^{-1}M_\odot$ out to $z=1$ drawn from the mass function for a given survey strategy (e.g., the chosen value of $A$, the survey area of $50$~deg$^2$). Having created the initial catalog of NFW halos each with ($c_{500},M'_{500}|M_{500},z$), we apply a selection cut of $M^\prime_{500} > M_\mathrm{lim}(z)$ to obtain an observable-selected sample of synthetic clusters. Finally, we create synthetic observations of cluster counts and weak-lensing profiles for the selected sample according to the procedure described in Section~\ref{subsec:obs}.

In this study, we select one realization of synthetic data (i.e., cluster counts and weak-lensing profiles) created with our fiducial model $F(\Omega_\mathrm{m}=0.286, \sigma_8=0.82, A=7.212)$ as an ``input'' dataset for our cosmological analysis. That is, the same forward simulator is used to create both the input dataset and forward simulations of synthetic data. This is an idealized assumption that is not necessarily true in reality. The recovered uncertainties are thus likely to be underestimated compared to real observations where one would explore a wide range of possible models.

In our cosmological forward simulations, it is assumed that we have perfect knowledge of the survey selection function except for the normalization $A$ of the $M_\mathrm{lim}(z)$ function and of the source redshift distribution for weak lensing. We also assume that weak lensing mass measurements are unbiased. As a result, we have three parameters for modeling our cluster observables (see Section~\ref{subsec:obs}), namely, $(\Omega_\mathrm{m}, \sigma_8, A)$. In our cosmological forward inference, we adopt the following uniform priors:
$\Omega_\mathrm{m}=1-\Omega_\Lambda\in [0.1, 0.5]$,
$\sigma_8\in[0.5, 1.0]$, and
$A\in [7.0, 7.5]$.

\begin{figure}[htbp]
 \centering
 \includegraphics[width=0.48\textwidth,clip]{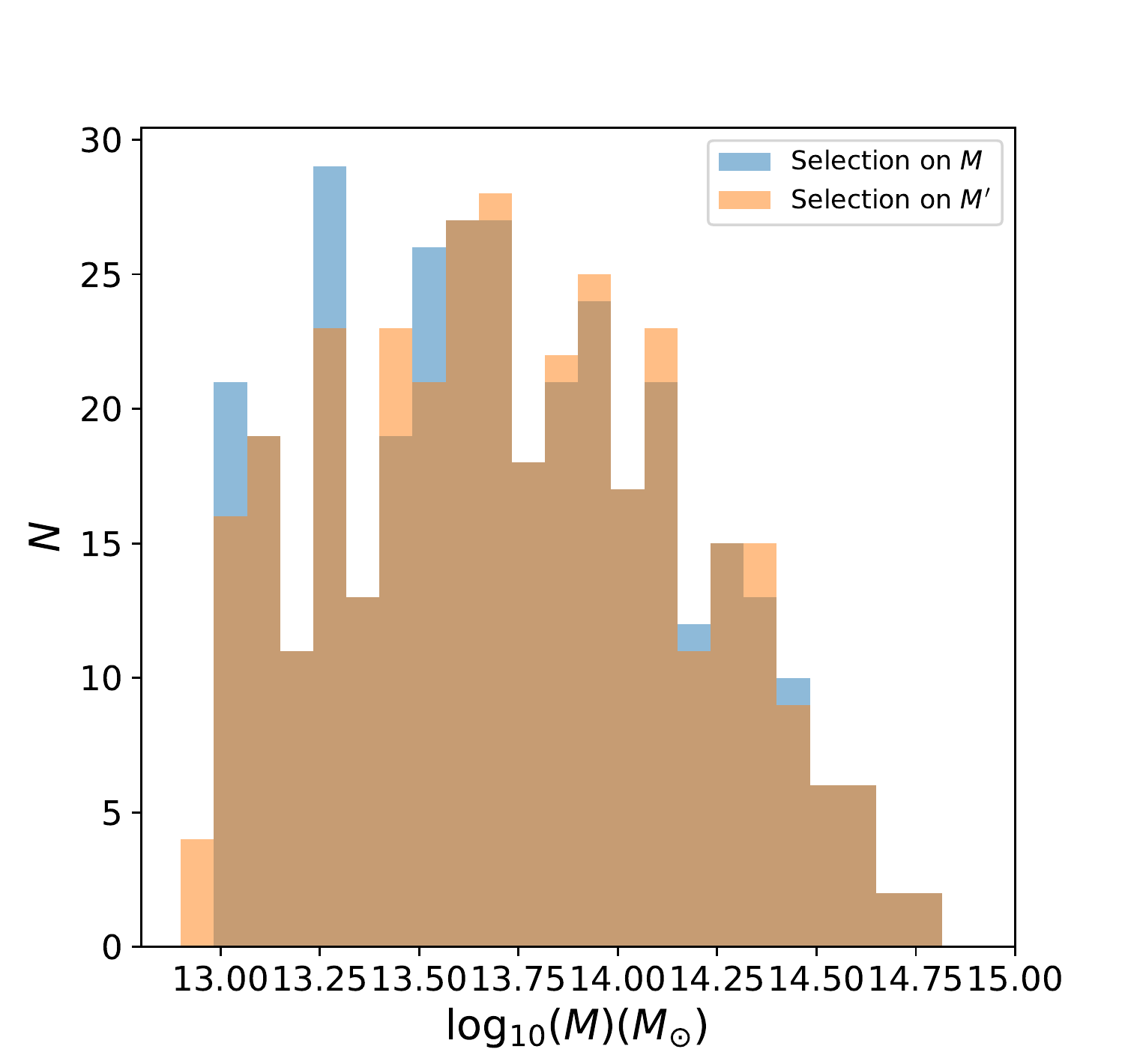}
 \caption{Histogram distribution of detected galaxy clusters as a function of true halo mass $M_{500}$. The blue (orange) histogram represents the cluster sample when the selection is applied on the true (scattered) halo mass. The cluster sample defined by the scattered observable $M'_{500}$ includes up-scattered low-mass halos below the minimum mass threshold $M_\mathrm{min}(z)$.}
 \label{fig:hist_M}
\end{figure}

\begin{figure}[htbp]
 \centering
 \includegraphics[width=0.48\textwidth,clip]{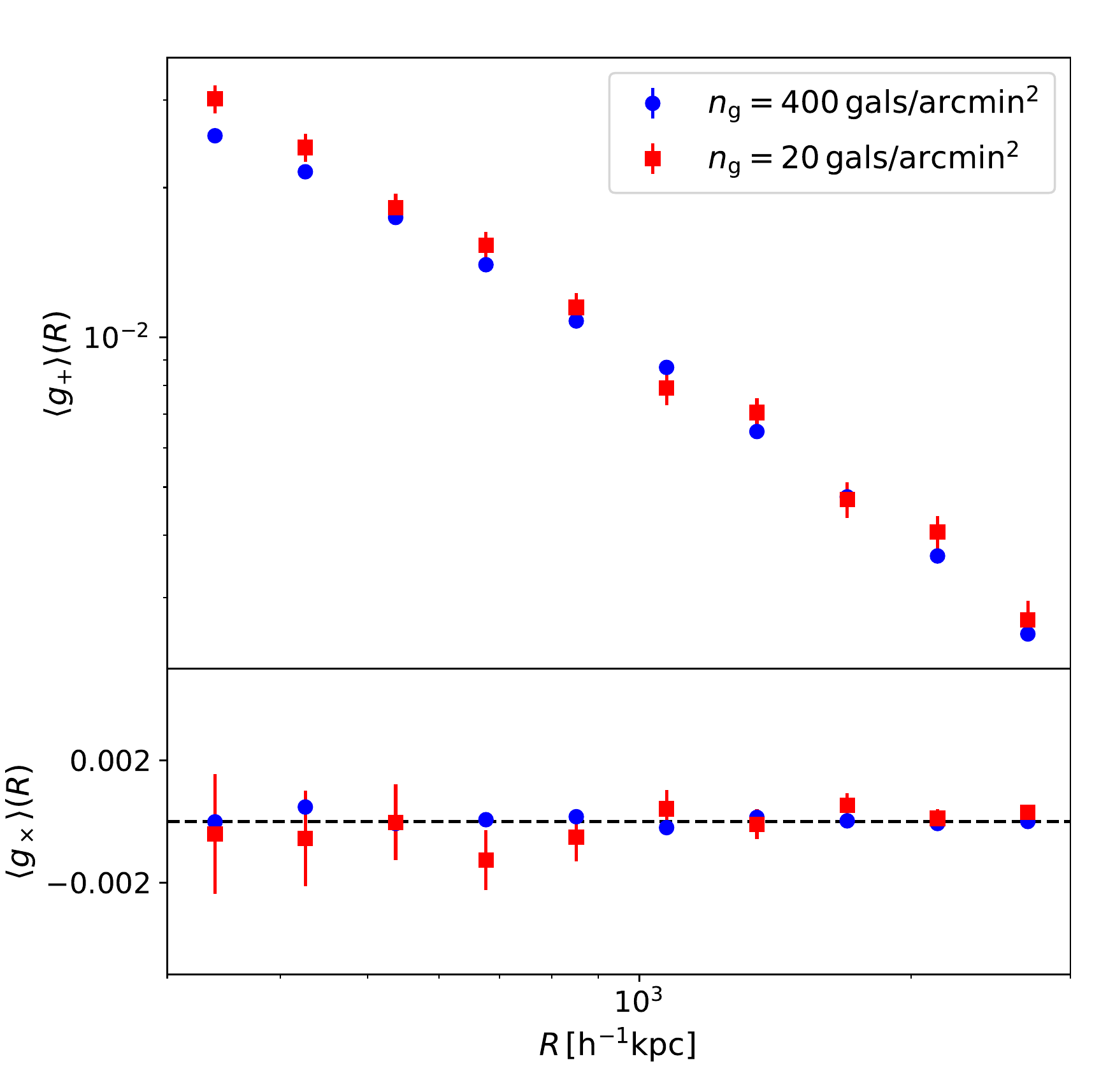}
 \caption{Azimuthally averaged reduced shear profiles of the stacked cluster samples derived from synthetic weak-lensing observations created with our fiducial model $F(\Omega_\mathrm{m}=0.286, \sigma_8=0.82, A=7.212)$. The upper (lower) panel shows the reduced tangential (cross) shear profile $\langle g_{+}\rangle(R)$ ($\langle g_{\times}\rangle(R)$) as a function of comoving cluster-centric radius $R$. The red squares with error bars show the results with $n_\mathrm{g}=20$~galaxies~arcmin$^{-2}$, while the blue circles with error bars show the results with $n_\mathrm{g}=400$~galaxies~arcmin$^{-2}$. We note that the two synthetic surveys detect different numbers of clusters (Table~\ref{tab:WL-SN}) corresponding to different realizations of scattered mass observables, with different masses of individual clusters.}
 \label{fig:gt_profile}
\end{figure}

\begin{deluxetable}{lccc}
\tablecolumns{4}
\tablewidth{0pt}
\tabletypesize{\scriptsize}
\tablecaption{\label{tab:WL-SN}
Signal-to-Noise Ratio (S/N) of the Stacked Cluster Lensing Profile}
\tablehead{
    \colhead{Survey sensitivity} & 
    \colhead{$N_\mathrm{cl}$\tablenotemark{a}}    &
    \colhead{$\mathrm{S/N}_+$\tablenotemark{b}}   & 
    \colhead{$\mathrm{S/N}_\times$\tablenotemark{c}}
}
\startdata
$n_\mathrm{g}=20$~arcmin$^{-2}$  & 325 & 44.2   & 2.49 \\
$n_\mathrm{g}=400$~arcmin$^{-2}$ & 336 &  191   & 2.80
\enddata
\tablenotetext{a}{Number of detected clusters for the particular realization of synthetic survey data.}
\tablenotetext{b}{S/N estimated from the stacked reduced tangential shear profile $\langle g_+\rangle(R)$.}
\tablenotetext{c}{S/N estimated from the stacked reduced cross shear profile $\langle g_\times\rangle(R)$.}
\end{deluxetable}

Figure~\ref{fig:hist_M} shows the number counts of detected galaxy cluster as a function of their true halo mass $M_{500}$ for one particular realization of synthetic observations to illustrate the impact of Eddington bias due to scatter in the observable--mass relation. The blue histogram represents the cluster sample when the selection (Equation~\ref{eq:mmin}) is applied on the true halo mass $M_{500}$, while the orange histogram represents the sample when the selection is applied on the scattered mass observable $M'_{500}$. 
%The former and the latter samples contain $N_\mathrm{cl}=328$ and $336$ selected clusters, respectively.
The cluster sample defined by the scattered mass observable $M'_{500}$ includes up-scattered low-mass halos below the minimum mass threshold $M_\mathrm{min}(z)$.  We note that the selection on the true mass $M_{500}$ is for illustration purposes only.
 
In Figure~\ref{fig:gt_profile}, we present the stacked reduced shear profiles $\langle g_+\rangle(R)$ and $\langle g_\times\rangle(R)$ derived from a synthetic weak-lensing dataset created with our simulator with our fiducial model, $F(\Omega_\mathrm{m}=0.286, \sigma_8=0.82$, $A=7.212)$. It should be noted that these two synthetic surveys detect different numbers of clusters (Table~\ref{tab:WL-SN}) corresponding to different realizations of intrinsic scatter, with different masses of individual clusters. The signal-to-noise ratios (S/N) of the stacked lensing profiles (see Figure~\ref{fig:gt_profile}) are listed in  Table~\ref{tab:WL-SN}.\footnote{To calculate the S/N, we use the conventional quadratic estimator defined with diagonal shape errors (see Equation~(114) of \citealt{Umetsu2020rev}).}

\subsection{ABC Inference} 
\label{sec: abc}

\begin{figure*}[htbp]
 \centering
 \includegraphics[width=0.85\textwidth,clip]{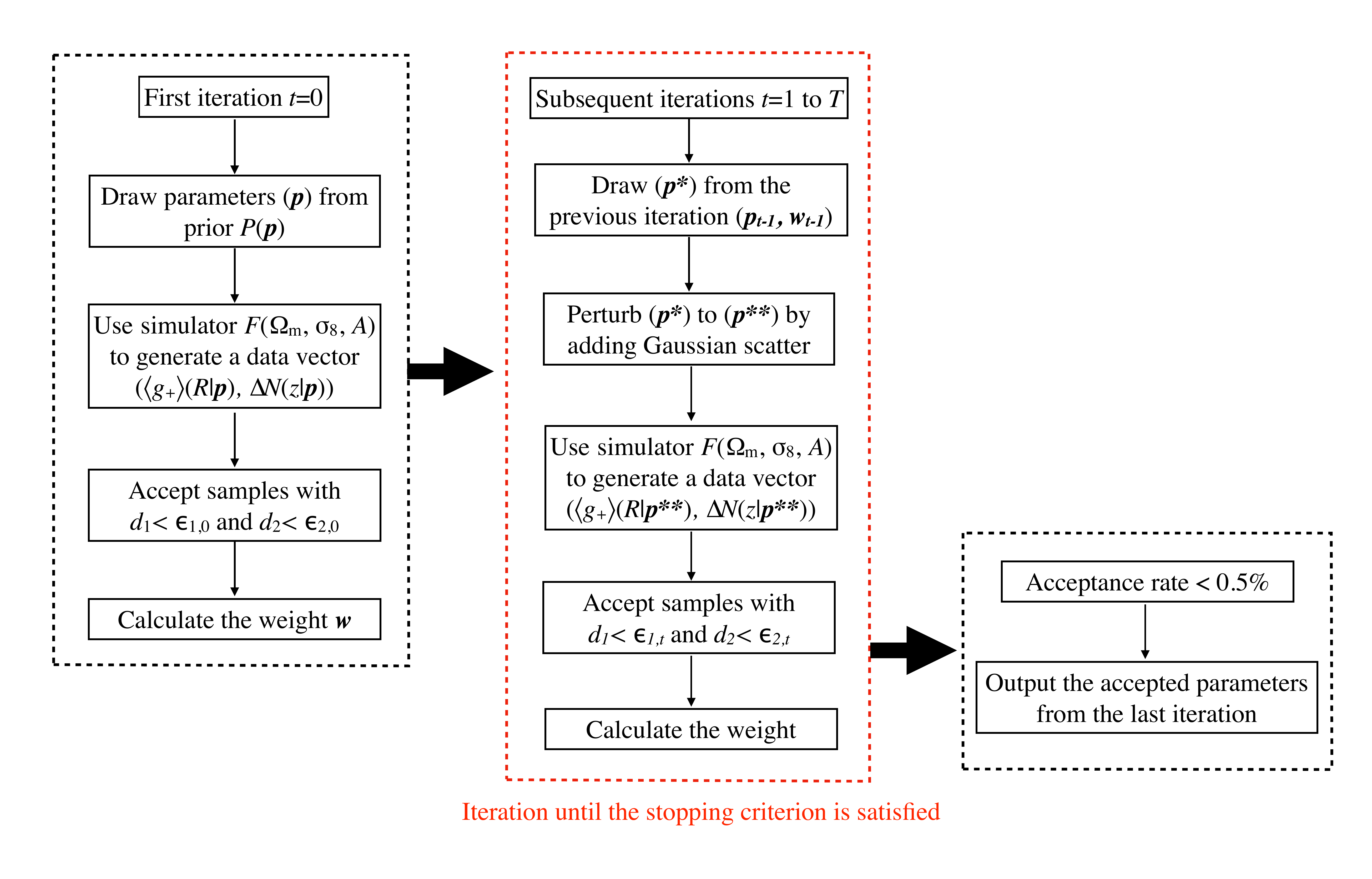}
 \caption{Schematic diagram illustrating the ABC Population Monte Carlo algorithm.}
 \label{fig:abc_flowchart}
\end{figure*}

Approximate Bayesian Computation (ABC) constitutes a family of likelihood-free inference methods suitable for statistical problems with intractable likelihoods, but where fast model evaluations with simulations are possible. The main advantage of ABC inference is that one can implement complex physical processes and instrumental effects into a simulation-based model, which is generally more straightforward compared to incorporating these effects in a likelihood function. Consequently, ABC has been widely applied in various areas of astrophysics and cosmology \citep[e.g.][]{2012LNS...902....3S,2012MNRAS.425...44C,2013ApJ...764..116W,2014A&A...569A..13R,2015A&A...583A..70L,abcpmc,2016arXiv161103087J,2017MNRAS.469.2791H,2018ApJ...855..106D,2018JCAP...02..042K,2020arXiv201013221H,2020JCAP...09..048T,2021arXiv210602651T}.

In the rejection ABC algorithm \citep{10.2307/2240995}, a synthetic data vector is generated from a forward simulator, given a set of input parameters (\bp) drawn from the prior distribution, $P(\bp)$.  A predefined distance metric measures the similarity between observed and simulated data. Parameters are accepted only if the synthetic data vector is within a user-specified threshold ($\epsilon$) from the observed data vector. The accepted parameters will then form a set of approximated posterior samples. As the thresholds decrease toward zero ($\epsilon\rightarrow0$), the ABC-derived posterior will tend to approach the true posterior distribution. 

During the optimisation process of rejection-based ABC, it is generally inefficient to propose parameters randomly drawn from an uninformative prior, because many simulations may be rejected. Therefore, variants of likelihood-free rejection algorithms, such as Population Monte Carlo ABC \citep[PMC;][]{2008arXiv0805.2256B,2015A&C....13....1I,abcpmc} and Sequential Monte Carlo ABC \citep[SMC;][]{SMC-2006,SMC-2009}, improve upon this situation by drawing parameters from an adaptive proposal distribution that identifies a more relevant portion of the parameter space. These advanced algorithms start from the prior distribution and converge to an approximate posterior by sampling parameters for a sequence of gradually decreasing thresholds ($\epsilon$). 

In this work, we use the \textsc{abc-pmc} package \citep{abcpmc} to perform our ABC analysis. We first define a distance metric $d_1$ for cluster weak-lensing observations using the stacked lensing observable as
\begin{equation}
\label{eq:d1}
    d_1=\sum_{i=1}^{N_\mathrm{bin}}\left[\langle g_{+}^{\mathrm{obs}}\rangle(R_i) -\langle g_{+}^{\mathrm{sim}}\rangle(R_i|\bp)\right]^2
\end{equation}
where $i$ runs over all radial bins, $\langle g_{+}^{\mathrm{obs}}\rangle(R_i)$ is the stacked shear measurement in the $i$th bin, 
\begin{equation}
\label{eq:gtobs}
\langle g_{+}^{\mathrm{obs}}\rangle(R_i) = \frac{1}{N_\mathrm{cl}}\sum_{m=1}^{N_\mathrm{cl}} g_{+,m}^{\mathrm{obs}}(R_i),
\end{equation}
and $\langle g_{+}^{\mathrm{sim}}\rangle(R_i|\bp)$ is a simulated realization given a set of model parameters $\bp$,
\begin{equation}
\label{eq:gtmod}
\langle g_{+}^{\mathrm{sim}}\rangle(R_i|\bp) = \frac{1}{N_\mathrm{cl}}\sum_{m=1}^{N_\mathrm{cl}} g_{+,m}^{\mathrm{sim}}(R_i|\bp).
\end{equation}
It should be noted that $\langle g_{+}^{\mathrm{sim}}\rangle(R_i|\bp)$ includes a realization of observational noise and, in general, a statistical fluctuation of the signal. The theoretical expectation of Equation~(\ref{eq:gtmod}) is given by Equation~(\ref{eq:stackedgt}).

Next, we define a distance metric for the cluster abundance as
\begin{equation}
\label{eq:d2}
    d_2=\sum_{k=1}^{N_z}\left[\Delta N^\mathrm{obs}(z_k)-\Delta N^\mathrm{sim}(z_k|\bp)\right]^2,
\end{equation}
where $k$ runs over all redshift bins ($1,2,\dots,N_z)$, $N_z$ is the number of redshift bins, $\Delta N^\mathrm{obs}(z_k)$ is the observed cluster counts in the $k$th bin, and $\Delta N^\mathrm{sim}(z_k|\bp)$ is a simulated realization of cluster counts given a set of model parameters $\bp$. In this work, we set $N_z=20$. We note that both $\Delta N^\mathrm{obs}(z_k)$ and $\Delta N^\mathrm{sim}(z_k|\bp)$ include statistical fluctuations from the intrinsic scatter in the observable--mass relation and the resulting effect of Eddington bias.

We define $\epsilon_1$ and $\epsilon_2$ to be the thresholds for the two distance metrics $d_1$ and $d_2$, respectively. A set of model parameters $\bp$ is accepted only when $d_1<\epsilon_1$ and $d_2<\epsilon_2$. The initial thresholds $\epsilon_{1,0}$ and $\epsilon_{2,0}$ are set to 1.0 and 2000.0, respectively. Following \citet{abcpmc}, we use an adaptive choice of the threshold such that the threshold for each distance metric ($d_1$ or $d_2$) is set to the 75th percentile of the accepted distances from the previous iteration. In this way, the thresholds will be automatically reduced during the iterative process.

An illustrative schematic of the \textsc{abc-pmc} algorithm is shown in Figure~\ref{fig:abc_flowchart}. In \textsc{abc-pmc}, the threshold $\epsilon$ becomes smaller and thus more realizations are rejected as the iteration proceeds. Thus, to have a fixed number of accepted samples (e.g., $10^3$ accepted samples in this work), we need to increase the number of realizations in subsequent iteration steps. For example, for the first iteration step with an acceptance rate $\sim 60\percent$, we need to generate $N=10^3/0.6\sim 1700$ realizations; while for the last iteration step with an acceptance rate of $\sim 0.5\percent$, we need to generate $N\sim 2\times 10^{5}$ realizations. 

Each set of parameters $(\Omega_\mathrm{m},\sigma_8,A)$ (hereafter referred to as a particle) sampled by the algorithm is assigned a weight $w$, which determines the probability of the particle being drawn in the next iteration. In the first iteration step, each particle has an equal weight, $w=1/N$, with $N$ the number of sampled particles (or the number of realizations). In subsequent iteration steps, the algorithm randomly samples particles from the parameter space by accounting for the probabilities, or the assigned weights. New weights are assigned to newly sampled particles. Since the weights determine the regions of parameter space to be efficiently explored, the weighting scheme is crucial for the efficiency of the algorithm. We also refer the reader to \citet{abcpmc} for more details. 

Taking into account both the finite computational resources available and the selection efficiency of the algorithm \citep{2019arXiv190701505S}, we define a stopping criterion such that the acceptance rate reaches a fixed threshold value of $0.5\percent$. In practice, when $\epsilon$ approaches small values, the approximated posterior begins to stabilize. A continued reduction of $\epsilon$ does not improve the accuracy of the inferred posterior significantly but results in a low acceptance rate \citep{2015A&C....13....1I,2015A&A...583A..70L,abcpmc}. For a further lower acceptance rate ($<0.5\percent$) corresponding to an even smaller threshold, the sampling process will become increasingly inefficient, so that a large fraction of the computational effort may be wasted.

\subsection{DELFI Inference}

As we discussed in Section~\ref{sec: abc}, it is computationally intensive to obtain a good posterior approximation (i.e., small enough $\epsilon$) using an ABC algorithm. Even using an advanced sampling algorithm such as \textsc{abc-pmc}, ABC methods suffer from the problem of vanishingly small acceptance rates when the threshold $\epsilon$ approaches zero \citep{2018MNRAS.477.2874A}. As a result, ABC requires an expensively large number of simulations. 
 
To overcome this problem of rejection-based ABC, we employ an $\epsilon$-free approach known as DELFI \citep[Density Estimate Likelihood-Free inference;][]{2016arXiv160506376P,2017arXiv171101861L,2018arXiv180507226P,2019MNRAS.488.4440A} as an alternative to the \textsc{abc-pmc} method. DELFI is a likelihood-free density-estimation approach to learn the sampling distribution of data as a function of the model parameters using neural density estimators.

In this study, we use the \textsc{pydelfi} package \footnote{https://github.com/justinalsing/pydelfi} \citep{2019MNRAS.488.4440A} to infer the posterior distribution. Recently, \textsc{pydelfi} has been used to study several inference problems \citep[e.g.][]{2019PhRvD.100b3519T,2021arXiv210503344Z,2021MNRAS.501..954J,2021arXiv210314378D,2021arXiv210402728G}. 
The algorithm uses simulations to learn the conditional density function $p(\bt|\bp)$, where $\bt$ represents a set of ``data summaries'' and $\bp$ represents a set of model parameters. The likelihood is then evaluated for a given observed data vector $\bt_o$ as $p(\bt_o|\bp)$. Multiplying this by the prior leads to the posterior $p(\bp|\bt_o)\propto p(\bt_o|\bp)\times p(\bp)$. 

The inference procedure of \textsc{pydelfi} is briefly summarised as follows: We first sample an initial set of parameters $\bp$ from the prior and create synthetic data summaries $\bt$ (a data vector containing summary statistics) using forward simulations. In our analysis, data summaries comprise the stacked cluster lensing profile and the cluster number counts in redshift bins, ($\langle g_{+}\rangle(R)$, $\Delta N(z)$). These data summaries are identical to the ones used in the ABC approach. Although \textsc{pydelfi} implements advanced data compression schemes to obtain a small number of informative data summaries, we use the same set of data summaries to achieve a direct comparison to the ABC approach.

DELFI uses flexible neural density estimators to learn the sampling distribution of data in the parameter space from a set of simulated data--parameter pairs ($\bt,\bp$).
\textsc{pydelfi} implements an active learning scheme with the sequential neural likelihood \citep[SNL;][]{2018arXiv180507226P} algorithm, which allows neural density estimators to draw new simulations from a proposal density based on the current posterior approximation. This algorithm adaptively learns the most relevant regions of the parameter space to run new simulations, thus improving the posterior inference.
During the training process, the neural density estimators are trained to learn the weights of the neural network, $w$, by minimizing the (negative log) loss function defined as
\begin{equation}
    -\ln(U)=-\sum_{i}^{N_{\mathrm{data}}}\ln{P(\bt_{i}|\bp_{i},w)},
    \label{eq:loss}
\end{equation}
which is equivalent to the negative log-likelihood of the simulation data $(\bt, \bp)$.

Finally, this density estimation network derives a sample of parameters to constitute the posterior distribution. For details of the algorithm used in \textsc{pydelfi}, we refer the reader to \cite{2016arXiv160506376P}, \cite{2017arXiv171101861L}, \cite{2018arXiv180507226P} and \cite{2019MNRAS.488.4440A}.

\section{Cosmological Parameter Inference}
\label{sec:results}

In this section, we present and discuss the results of likelihood-free cosmological inference using our forward simulator. In Appendix~\ref{sec:toymodel}, we present two toy models for weak-lensing mass calibration to demonstrate the potential and performance of likelihood-free methods along with the conventional maximum-likelihood approach (see Section~\ref{sec:conclusion} for a summary of the findings).

\subsection{Results and Discussion}
\label{sec:cosmo}

%%% Table 4
\begin{deluxetable*}{lcccc|cccc}
\tablecolumns{9}
\tablewidth{0pt}
\tabletypesize{\scriptsize}
\tablecaption{\label{tab:cosmoresult}
Posterior Summaries for Cosmological Parameter Inference}
\tablehead{
 \colhead{Survey sensitivity}  & 
 \multicolumn{4}{c}{ABC-PMC} &  
 \multicolumn{4}{c}{PYDELFI}   \\   
 \cline{2-9} 
 \colhead{} &
 \multicolumn{1}{c}{$\Omega_\mathrm{m}$ } &
 \multicolumn{1}{c}{$\sigma_8$} &
 \multicolumn{1}{c}{$A$} &
 \multicolumn{1}{c}{$S_8$} &
 \multicolumn{1}{c}{$\Omega_\mathrm{m}$ } &
 \multicolumn{1}{c}{$\sigma_8$} &
 \multicolumn{1}{c}{$A$} &
 \multicolumn{1}{c}{$S_8$} 
}
\startdata
$n_\mathrm{g}=20$~arcmin$^{-2}$ & $0.269 \pm 0.035$  & $0.866 \pm 0.034$ & $7.236 \pm 0.036$ & $0.836\pm 0.032$ & $0.256\pm 0.013$ & $0.847 \pm 0.024$ & $7.213 \pm 0.012$ & $0.810\pm 0.019$\\
$n_\mathrm{g}=400$~arcmin$^{-2}$ & $0.264\pm 0.023$ & $0.825\pm 0.025$
 & $7.208\pm 0.021$ & $0.793\pm 0.018$ & $0.267\pm 0.013$ &  $0.812\pm 0.020$ & $7.209 \pm 0.009$ & $0.784\pm 0.013$
\enddata
\end{deluxetable*}

\begin{figure}[htbp]
 \centering
 \includegraphics[width=0.48\textwidth,clip]{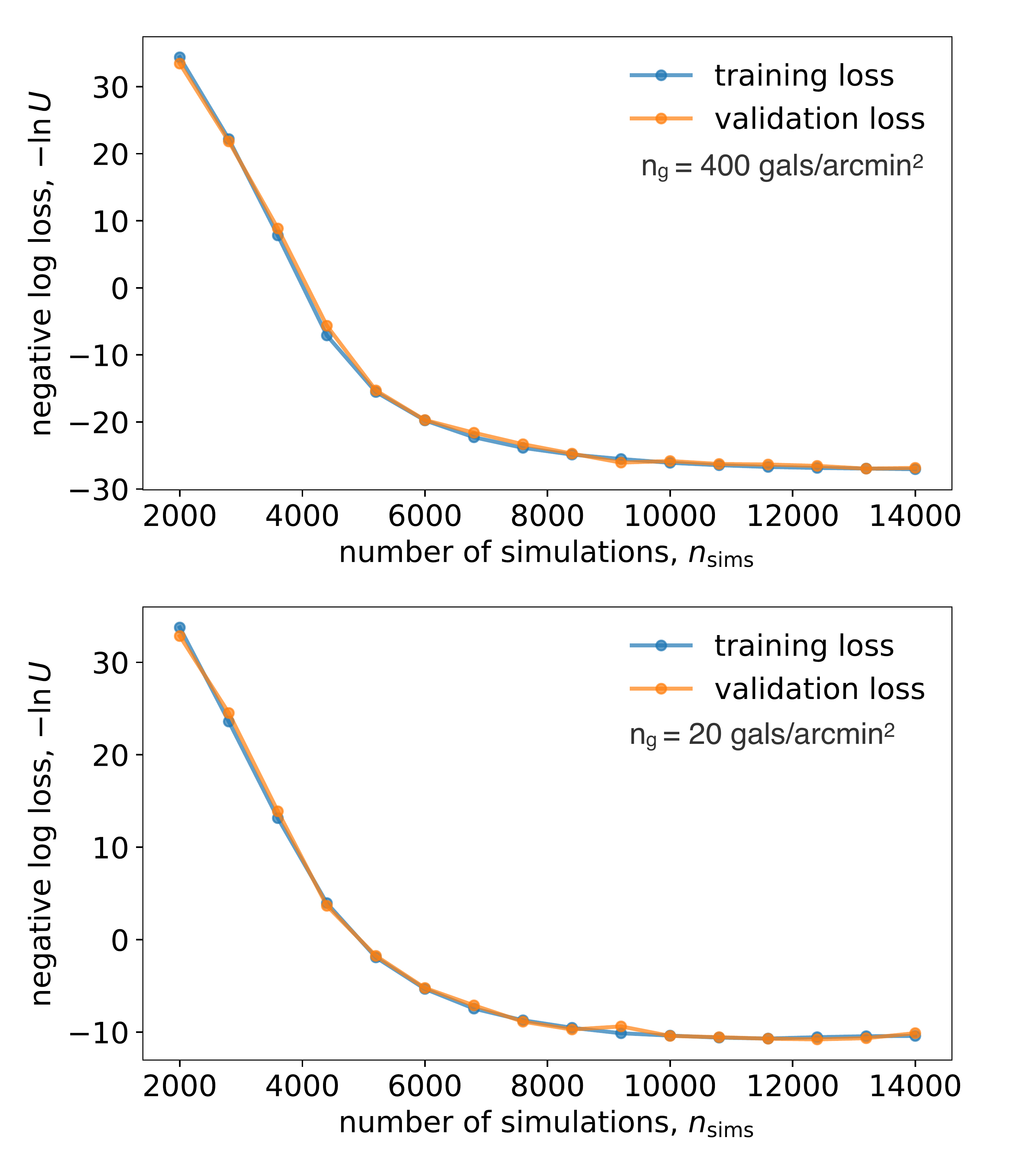}
 \caption{Minimum value of the loss function as a function of the total cumulative number of simulations for cosmological inference using \textsc{pydelfi}. The results are shown separately for two different survey depths, $n_\mathrm{g}=400$~galaxies~arcmin$^{-2}$ (upper panel) and $n_\mathrm{g}=20$~galaxies~arcmin$^{-2}$ (lower panel). In each panel, the values are shown for both training (blue) and validation (orange) samples.}
 \label{fig:delfiloss}
\end{figure}

\begin{figure}[htbp]
 \centering
 \includegraphics[width=0.48\textwidth,clip]{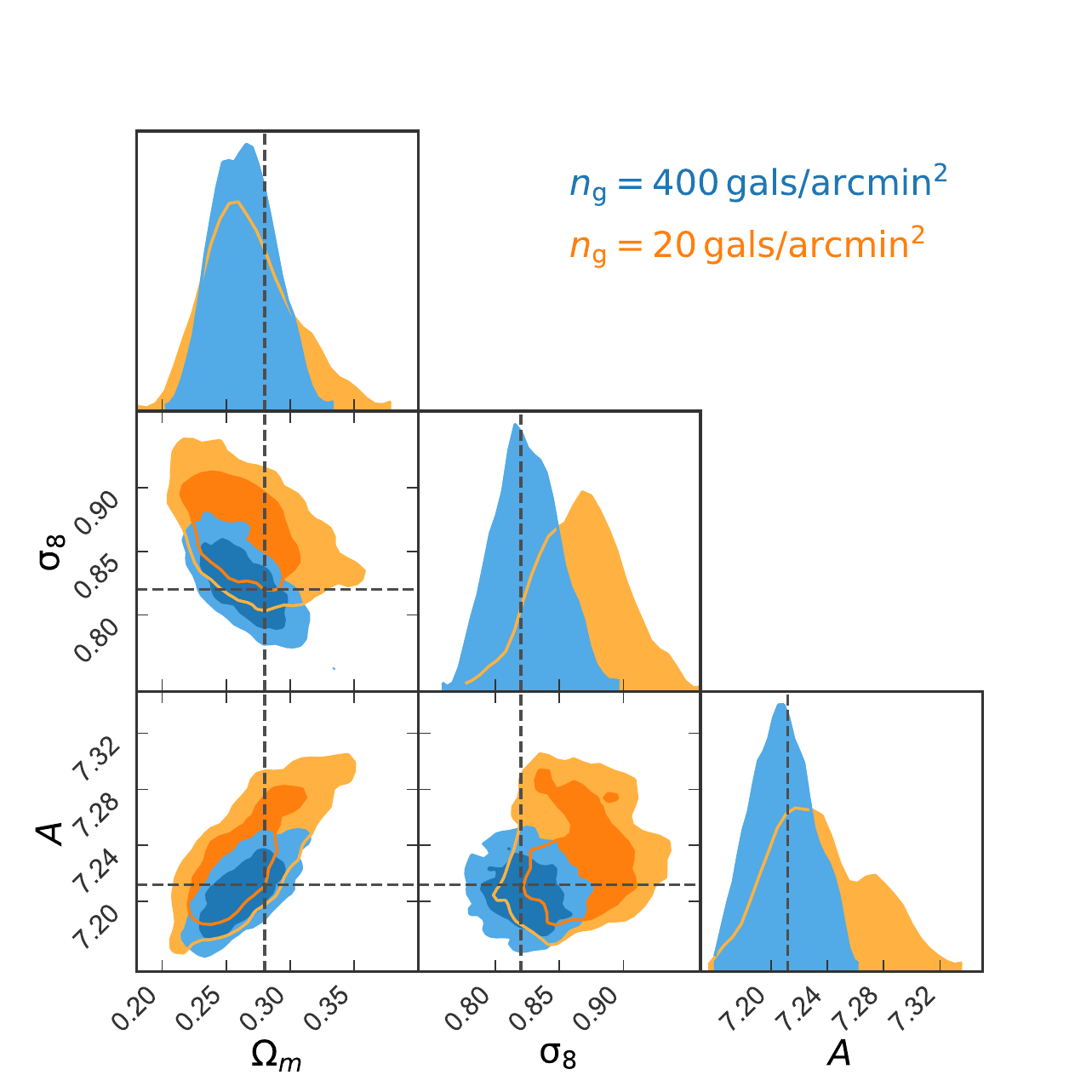}
 \caption{Parameter constraints showing marginalized one-dimensional (histograms) and two-dimensional (68\percent\ and 95\percent\ confidence level contours) posterior distributions obtained with the \textsc{abc-pmc} approach. For each parameter, the black dashed line indicates the fiducial value assumed in this study. A summary of the marginalized posterior constraints on the parameters ($\Omega_\mathrm{m}, \sigma_8, A$) is given in Table~\ref{tab:cosmoresult}.}
 \label{fig:abccosmo}
\end{figure}

\begin{figure}[htbp]
 \centering
 \includegraphics[width=0.48\textwidth,clip]{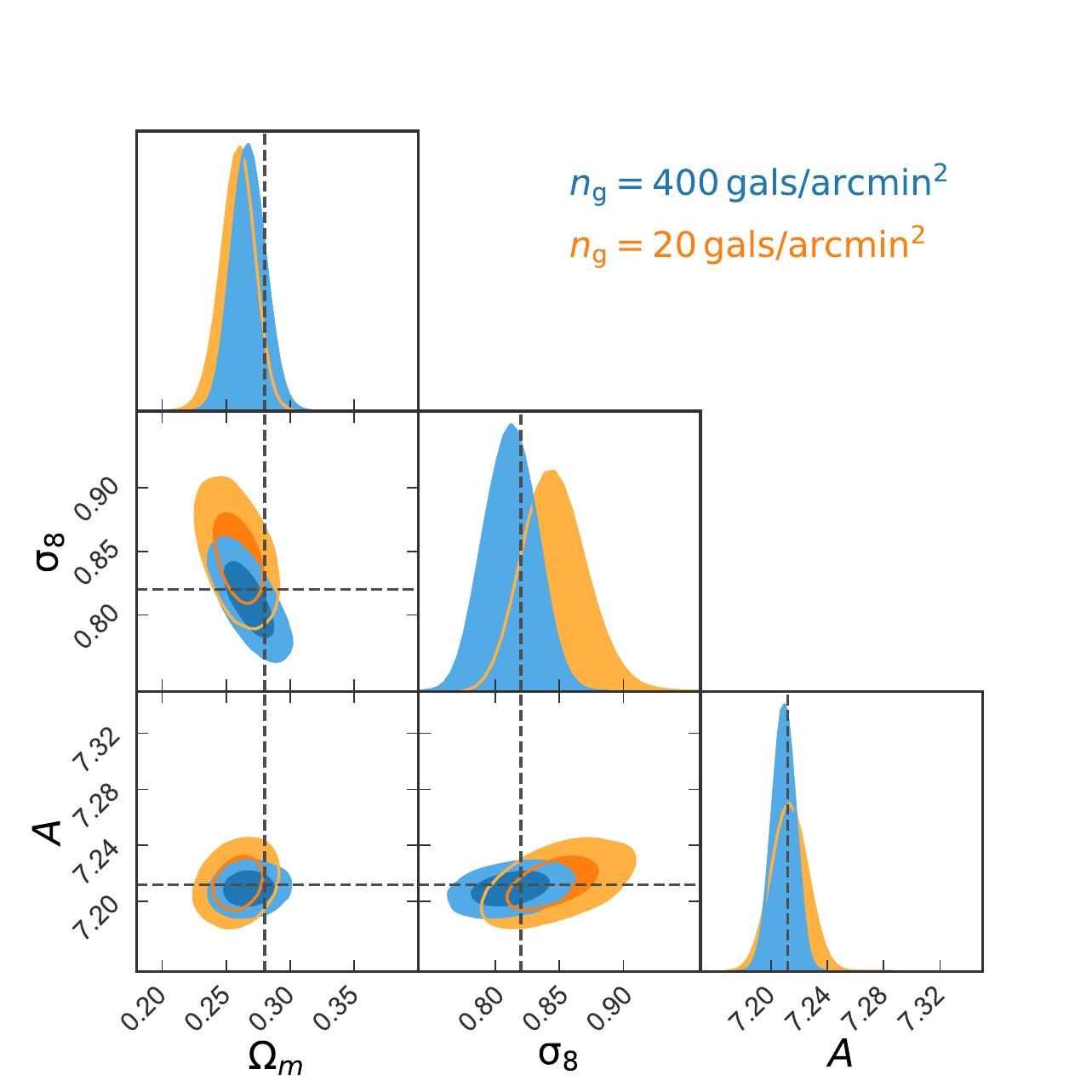}
 \caption{Same as in Figure~\ref{fig:abccosmo}, but obtained with the \textsc{pydelfi} approach. A summary of the marginalized posterior constraints on ($\Omega_\mathrm{m}, \sigma_8, A$) is given in Table~\ref{tab:cosmoresult}.}
 \label{fig:delficosmo}
\end{figure}

In this subsection, we perform cosmological parameter inference from a synthetic cluster survey using \textsc{abc-pmc} and \textsc{pydelfi}. We use our fiducial model $F(\Omega_\mathrm{m}=0.286, \sigma_8=0.82, A=7.212)$ described in Section~\ref{sec:simulation} to generate synthetic datasets. In both \textsc{abc-pmc} and \textsc{pydelfi} analyses, we use the stacked lensing profile $\langle g_{+}\rangle(R)$ and the cluster counts in redshift bins $\Delta N(z)$  (Section~\ref{sec:simulation}) as summary statistics, or data summaries. 

We run \textsc{abc-pmc} in a series of iterations, in each of which $10^3$ accepted samples are generated. We also use the same stopping criterion defined in Section~\ref{sec: abc}. With this setup, more than $\mathcal{O}(10^6)$ forward simulations are required.

To set up \textsc{pydelfi}, we implement an ensemble of six neural density estimators, including firstly one masked autoregressive flow with six masked autoencoders for density estimations each containing 2 hidden layers of 30 hidden units,  and secondly five mixture density networks  with $1, 2, \dots, 5$ Gaussian components respectively, with each mixture density network containing 2 hidden layers of 30 hidden units.  We have a total of 15 training steps with 2000 simulations for an initial training step. A batch-size of 800 in each training cycle is also given. A learning rate of $1\times10^{-5}$ is used, so that the minimum value of the loss function gradually decreases with the total cumulative number of simulations (see Figure~\ref{fig:delfiloss}).  

The resulting posterior distributions of the cosmological parameters obtained using \textsc{abc-pmc} and \textsc{pydelfi} are shown in Figures~\ref{fig:abccosmo} and \ref{fig:delficosmo}, respectively. In each figure, we show the results for two different survey depths, $n_\mathrm{g}=20$~galaxies~arcmin$^{-2}$ and $n_\mathrm{g}=400$~galaxies~arcmin$^{-2}$.
 Both likelihood-free methods provide unbiased and consistent posterior constraints, while \textsc{abc-pmc} recovers broader posteriors than \textsc{pydelfi}. 
 
\begin{figure}[htbp]
 \centering
 \includegraphics[width=0.48\textwidth,clip]{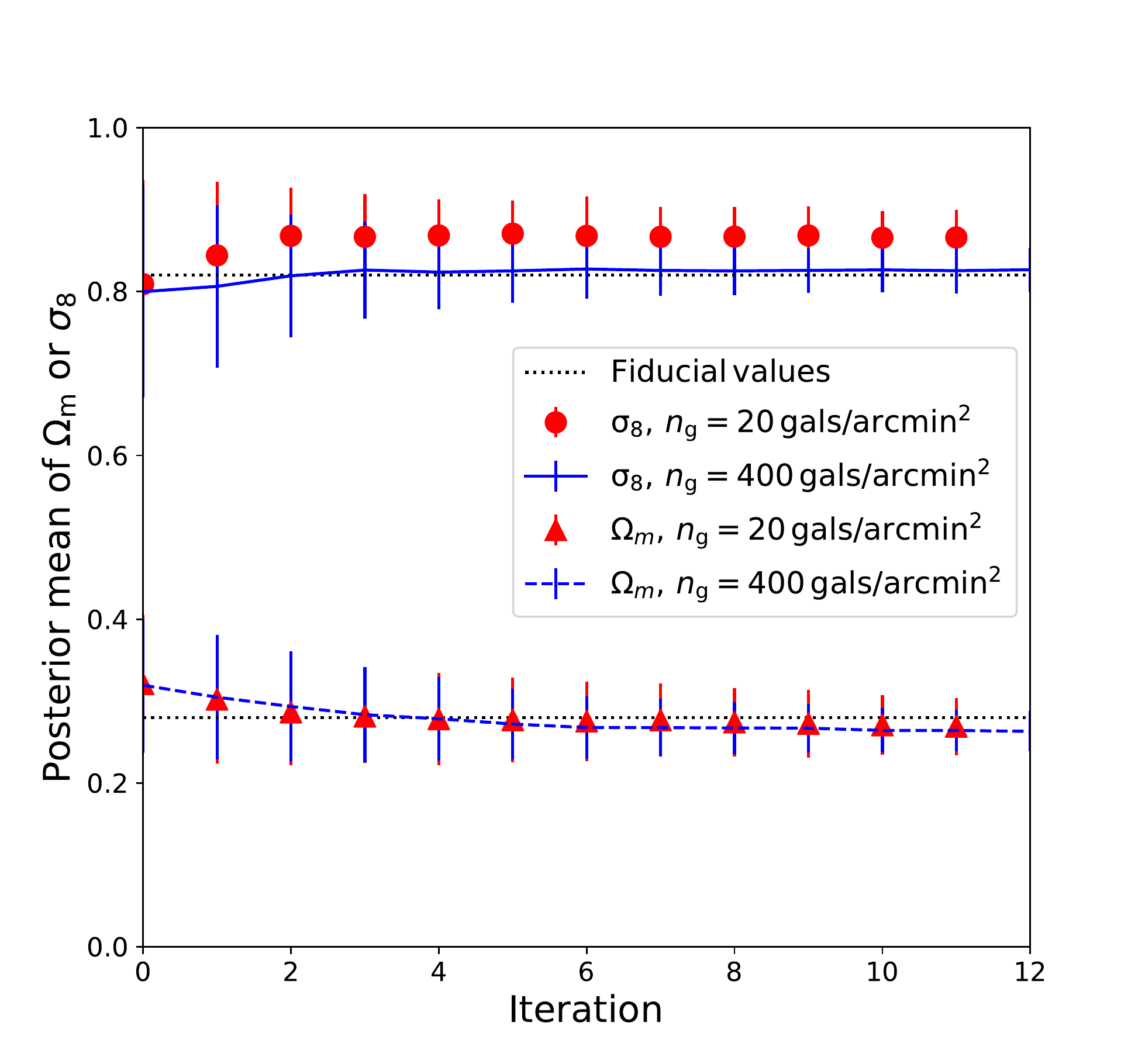}
 \caption{Convergence of the cosmological parameters inferred by the \textsc{abc-pmc} algorithm. Posterior means and errors of $\Omega_\mathrm{m}$ and $\sigma_8$ are shown in each iteration step for two different survey depths, $n_\mathrm{g}=20$~galaxies~arcmin$^{-2}$ (red: $\Omega_\mathrm{m}$ with triangles, $\sigma_8$ with circles) and $n_\mathrm{g}=400$~galaxies~arcmin$^{-2}$ (blue: $\Omega_\mathrm{m}$ with a dashed line, $\sigma_8$ with a solid line). The black dotted lines denote the input values of the parameters, $\Omega_\mathrm{m}=0.286$ and $\sigma_8=0.82$.}
 \label{fig:errorplot}
\end{figure}
 
With a sufficient amount of simulations, ABC produces a reasonable approximation to the posterior distribution, which is unbiased but broader than the true posterior.
 The more simulations we produce, the more accurate the approximated posterior will be, but at the expense of
increased computational run time. In Figure~\ref{fig:errorplot}, we demonstrate the convergence of the cosmological parameters inferred by the \textsc{abc-pmc} algorithm. The figure plots the posterior means and errors of $\Omega_\mathrm{m}$ and $\sigma_8$ in each iteration step for two different survey depths. The marginalized uncertainties of $\Omega_\mathrm{m}$ and $\sigma_8$ decrease gradually and their posterior means converge after a sufficient number of iterations.

Posterior summaries of the model parameters ($\Omega_\mathrm{m},\sigma_8,A$) are listed in Table~\ref{tab:cosmoresult}. In our cosmological inference, we find that \textsc{abc-pmc} gives more conservative errors than \textsc{pydelfi}. In particular, the uncertainty of $\Omega_\mathrm{m}$ obtained from \textsc{abc-pmc} is larger by a factor of 2--3 than that from \textsc{pydelfi}. Similarly, the uncertainty of $\sigma_8$ from \textsc{abc-pmc} is $30\percent$--$40\percent$ larger than that from \textsc{pydelfi}.  Since we analyze only one realization of synthetic data for each survey depth (as in the case of real experiments), it is expected that the posterior means are deviated from the ground truth (Figures~\ref{fig:abccosmo} and \ref{fig:delficosmo}; see also Appendix~\ref{sec:toymodel}).\footnote{We emphasize that such deviations in the inferred parameters are inevitable as long as there is only one realization of observations to be analyzed. If we were to analyze a large number of independent sets of synthetic observations and average them over, we expect to precisely recover the ground truth.}

In each survey depth, the posterior means inferred from \textsc{abc-pmc} and \textsc{pydelfi} are in agreement with each other, having the same direction of the shift, and they are consistent within the errors with the true value.
 
\begin{figure}[htbp]
 \centering
 \includegraphics[width=0.48\textwidth,clip]{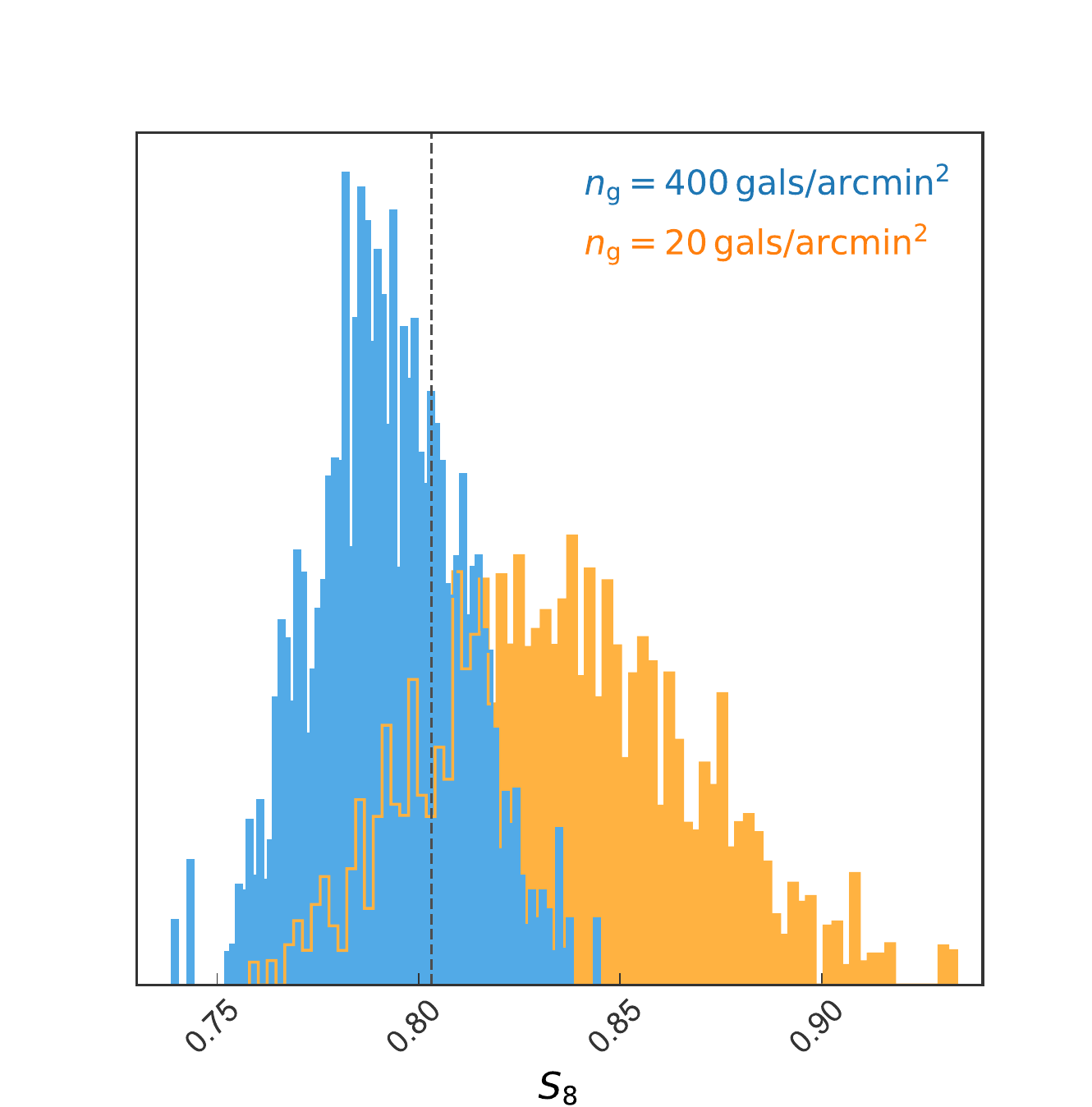}
 \caption{Marginalized posterior distribution of $S_8=\sigma_8(\Omega_\mathrm{m}/0.3)^{0.3}$ obtained with the \textsc{abc-pmc} approach. The blue and orange histograms show the results for $n_\mathrm{g}=400$ and $20$~galaxies~arcmin$^{-2}$, respectively. The vertical dashed line indicates the fiducial value assumed in this study. A summary of the marginalized posterior constraints on $S_8$ is given in Table~\ref{tab:cosmoresult}.}
 \label{fig:abc_s8}
\end{figure}

\begin{figure}[htbp]
 \centering
 \includegraphics[width=0.48\textwidth,clip]{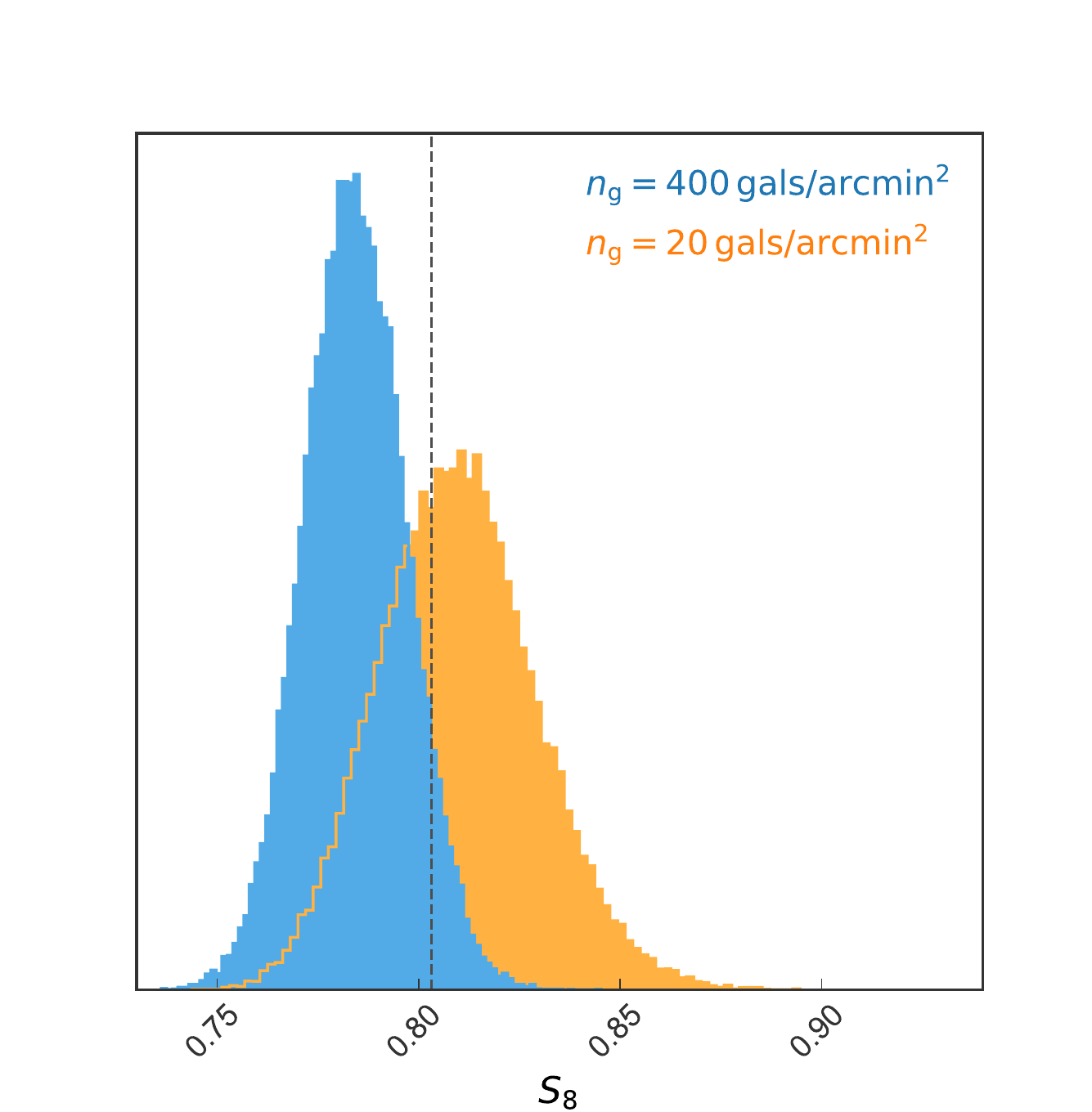}
 \caption{Same as in Figure~\ref{fig:abc_s8}, but obtained with the \textsc{pydelfi} approach. A summary of the marginalized posterior constraints on $S_8$ is given in Table~\ref{tab:cosmoresult}.}
 \label{fig:delfi_s8}
 \end{figure}

 In Figure~\ref{fig:abc_s8} and \ref{fig:delfi_s8}, we show the marginalized posterior distribution of $S_8=\sigma_8(\Omega_\mathrm{m}/0.3)^{0.3}$ obtained from the \textsc{abc-pmc} and \textsc{pydelfi} methods, respectively (see also Table~\ref{tab:cosmoresult}). For the case of $n_\mathrm{g}=400$~galaxies~arcmin$^{-2}$ with an idealized survey depth, the uncertainty in $S_8$ is largely dominated by the statistical fluctuation of the survey sample. The $\sigma_8$ parameter is sensitive to the mass calibration, so that the uncertainty in $S_8$ is increased substantially when the mean number density of background galaxies is decreased from $n_\mathrm{g}=400$~galaxies~arcmin$^{-2}$ to $20$~galaxies~arcmin$^{-2}$.

\subsection{Covariance Structure in Data}

The main advantage of the likelihood-free approach is its ability to implement complex physical processes, observational conditions, and instrumental effects into forward modeling. One of the key difficulties in the standard cosmological inference based on the Gaussian likelihood occurs in the derivation of the full covariance matrix. In likelihood-free methods, by contrast, all relevant statistical fluctuations due to observational noise and underlying cosmological/astrophysical signals are properly encoded in forward simulations.

%Figure~\ref{fig:correlation} shows the correlation matrix averaged over 20 realizations of cosmological forward simulations created with our fiducial model $F(\Omega_\mathrm{m}=0.28, \sigma_8=0.82, A=7.212$), 
%\begin{equation}
%    r_{ij}=\frac{\mathrm{Cov}(X_i,X_j)}{\sigma(X_i)\sigma(X_j)},
%\end{equation}
%where $X_i$ ($i=1,2,\dots,N_\mathrm{bin})$ denotes the reduced tangential shear around clusters in the $i$th radial bin, $\mathrm{Cov}(X_i, X_j)$ is the covariance  between $X_i$ and $X_j$, and $\sigma(X_i)$ is the standard deviation of $X_i$. The covariance matrix $\mathrm{Cov}(X_i, X_j)$ characterizes fluctuations of individual cluster profiles around their stacked mean profile. The results are shown separately for $n_\mathrm{g}=20$ and $400$~galaxies~arcmin$^{-2}$. 
%For the case with the highest shape noise (Ng=1 gal/arcmin$^2$), lensing signals are dominated by the random shape noise. Therefore, less correlation between different radial bins are seen. On the other hand,
%For the case of $n_\mathrm{g}=400$~galaxies~arcmin$^{-2}$, the covariance matrix is dominated by the variation of the lensing signal due to the spread in halo mass of the sample. Accordingly, a large amount of correlation between radial bins is detected. In contrast, for $n_\mathrm{g}=20$~galaxies~arcmin$^{-2}$, the covariance matrix is dominated by random shape noise, so that a much less amount of correlation is found.

In the likelihood-free approach, we need not model or quantify the covariance matrix a priori. Instead, as long as we properly account for and implement the relevant physics and observational effects in forward simulations, all statistical information will be fed into \textsc{abc-pmc} or \textsc{pydelfi}, which avoids complex derivation of the covariance matrix in a highly nonlinear and inherently complex problem.

%\begin{figure*}
%    \centering
%    \includegraphics[scale=0.28]{correlation_ng20_ng400_ave20.pdf}
%    \caption{Correlation coefficients of the covariance matrix for cluster weak-lensing profiles averaged over 20 realizations of synthetic weak-lensing observations. 
%    The results are shown for two different survey-depth parameters: $n_\mathrm{g}=20$~galaxies~arcmin$^{-2}$ (left) and $n_\mathrm{g}=400$~galaxies~arcmin$^{-2}$ (right).} 
%    \label{fig:correlation}
%\end{figure*}

\subsection{Modeling Assumptions and Current Limitations}
\label{subsec:assumptions}

In this subsection, we summarize the simplifying assumptions and limitations made in our current forward-modeling pipeline and discuss possible improvements to be made. 

First, we have assumed a spherical NFW description to model individual cluster halos. Collisionless cosmological $N$-body simulations predicted that cluster-scale dark matter halos are nonspherical and better described as triaxial halos with a preference for prolate shapes \citep[e.g.][]{2002ApJ...574..538J,2005ApJ...618....1H,Despali2017}. Moreover, the current modeling procedure neglects the intrinsic scatter in halo concentration at fixed halo mass. These effects will introduce substantial scatter in the projected cluster lensing signal at fixed halo mass \citep[a total of $\sim 20\percent$ scatter in the cluster lensing signal; see][]{Gruen2015,Umetsu2016}. Therefore, a more realistic halo description with triaxial NFW density profiles with a scattered $c$--$M$ relation \citep[e.g.][]{2018ApJ...860..126C} is expected to improve our cluster mass modeling for weak-lensing simulations. Nevertheless, we note that most recent lensing mass calibration studies for X-ray cluster surveys \citep{Umetsu2020hsc,Chiu2021} adopted an NFW halo description in their Bayesian population modeling. In these studies, the weak-lensing inferred mass is statistically calibrated using numerical simulations.

Second, our modeling focuses on the lensing signal produced by a single cluster halo, without including any contribution from subhalos or large-scale environments, i.e., the 2-halo term  \citep{2002PhR...372....1C}. The 2-halo term describes large-scale clustering properties of matter around dark matter halos, which contains crucial cosmological information. At smaller scales, other systematic effects that can affect the interpretation of observed cluster lensing profiles include cluster miscentering and residual contamination of the lensing signal by cluster members \citep[e.g.,][]{Chiu2021}. An implementation of such small- and large-scale modeling in our cosmological forward simulations will be a subject of future work.

Third, in this study, we have made various simplifications of background source and noise properties. In particular, we assumed perfect knowledge of the source redshift distribution and a constant background galaxy density $n_\mathrm{g}$ for all clusters out to $z=1$. Moreover, we neglected the effect of cosmic noise covariance on cluster lensing measurements as well as the intrinsic scatter in halo concentration. All these effects will act to reduce the statistical precision of weak-lensing mass calibration. Our future studies will include these realistic observational effects in our forward simulations.

Finally, this study has considered as summary statistics a single stack of the cluster lensing signal averaged over the full sample (Equation~\ref{eq:d1}). However, the redshift evolution of cluster density profiles and the geometric scaling of the lensing signal as a function of source redshift contain a wealth of cosmological information \citep[e.g.,][]{Taylor2007,Medezinski2011}, which we have not included in our cosmological inference. We will explore this possibility in our future work.

\subsection{Comparison with Observational Results from Cluster Surveys}

In this subsection, we first briefly summarize observational constraints on the cosmological parameters $\Omega_\mathrm{m}$ and $\sigma_8$ (or $S_8$) from recent cluster programs in the published literature.
\cite{2015MNRAS.446.2205M} obtained cosmological constraints using a sample of 50 high-mass X-ray clusters targeted by the Weighing the Giants program \citep{vonderLinden2014}. By combining cosmological information from X-ray observations with direct weak-lensing mass measurements, they obtain $S_8=\sigma_8(\Omega_\mathrm{m}/0.3)^{0.17} = 0.81 \pm 0.03$, or 
$\Omega_\mathrm{m} = 0.26 \pm 0.03$ and $\sigma_8= 0.83 \pm 0.04$. \cite{2016ApJ...832...95D} analyzed a sample of 377 clusters from the South Pole Telescope survey, finding $S_8=  \sigma_8(\Omega_\mathrm{m}/0.27)^{0.3}= 0.797\pm0.031$, or  $\Omega_\mathrm{m}=0.289 \pm 0.042$ and  $\sigma_8= 0.784 \pm 0.039$.  \cite{2017MNRAS.471.1370S} combined observational constraints on the mass function and gas mass fractions for 64 HIFLUGCS galaxy clusters to obtain $\Omega_\mathrm{m} = 0.30 \pm 0.01$ and $\sigma_8= 0.79 \pm 0.03$. \cite{2018A&A...620A..10P} analyzed the redshift distribution of 178 X-ray groups and clusters detected by the 50~deg$^2$ \textit{XMM}-XXL survey, finding $\Omega_\mathrm{m} = 0.316 \pm 0.060$ and $\sigma_8= 0.814 \pm 0.054$. 

Here we turn to discuss our inference results based on synthetic observations with $n_\mathrm{g}=20$~galaxies~arcmin$^{-2}$ by comparison to the cosmological constraints from cluster observations summarized above. For our \textsc{abc-pmc} inference, the uncertainties of the inferred parameters are comparable to these observational results. For our \textsc{pydelfi} inference, the uncertainties of the cosmological parameters are smaller than the observational ones. Our smaller uncertainties in $\Omega_\mathrm{m}$ are likely due in part to the small amount of scatter assumed for the observable--mass relation ($10\percent$ intrinsic scatter and no measurement uncertainty). Moreover, in our idealized setup, it is assumed that we have perfect knowledge of the selection function and the observable--mass relation out to $z= 1$, which helps break the parameter degeneracy between $\Omega_\mathrm{m}$ and $\sigma_8$ and thus reduce the uncertainty on $\Omega_\mathrm{m}$. 

In contrast, the size of the uncertainty in $\sigma_8$ is closer to the observational results. However, we reiterate that as a consequence of various simplifications made in our simulations (see Section~\ref{subsec:assumptions}), the uncertainty in $\sigma_8$ is expected to be underestimated.
%However,  direct comparisons to the observational dataset are complicated since we simplify our halo model in some context, thus a tighter constraints are obtained.

\section{Conclusions and Summary}
\label{sec:conclusion}

In this paper, we have explored the potential of likelihood-free inference of cosmological parameters from the redshift evolution of the cluster abundance combined with weak-lensing mass calibration. Likelihood-free inference provides an alternative way to perform Bayesian analysis using forward simulations only. The main advantage of likelihood-free methods is its ability to incorporate complex physical and observational effects in forward simulations. We employed two complementary likelihood-free methods, namely Approximate Bayesian Computation (ABC) and Density-Estimation Likelihood-Free Inference (DELFI), to develop an analysis procedure for inference of the cosmological parameters $(\Omega_\mathrm{m},\sigma_8)$ and the mass scale of the survey sample ($A$). These likelihood-free approaches allow us to bypass the need for a direct evaluation of the likelihood using forward simulations. In this study, we used two publicly available software packages, \textsc{abcpmc} \citep{abcpmc} and \textsc{pydelfi} \citep{2019MNRAS.488.4440A}, which implement the ABC and DELFI algorithms respectively.

To demonstrate the utility of likelihood-free methods, we presented in Appendix~\ref{sec:toymodel} two simplified toy models of weak-lensing mass calibration, where we neglect the scatter in the observable--mass relation and fix the number of selected clusters. In addition to the ABC and DELFI methods, we also employed a conventional maximum-likelihood (ML) method based on a single-mass-bin NFW fit to the stacked lensing signal. We find that for a simple problem (Toy model~\RN{1}) in which the underlying likelihood is well described by a Gaussian, both likelihood-free and conventional ML approaches obtain an unbiased recovery of the model parameters. For a more complex problem (Toy model~\RN{2}), forward-modeling approaches can properly account for all relevant statistical effects, which are encoded in the resulting posterior distributions. 

In general, a full description of complicated physical and observational effects is difficult to implement in the likelihood function. The use of the covariance matrix constructed from numerical simulations has to rely on the Gaussian likelihood assumption. Compared to the conventional Bayesian analysis, forward-modelling methods provide a more flexible framework that allows us to incorporate complex processes, which improves upon the completeness and accuracy of parameter inference.

Assuming an \textit{eROSITA}-like selection function (Figure~\ref{fig:selection}; \citealt{2012MNRAS.422...44P}) and a $10\percent$ scatter in the observable--mass relation in a flat $\Lambda$CDM cosmology ($\Omega_\mathrm{m}=0.286,\sigma_8=0.82)$, we create with our simulator a synthetic dataset of observable-selected NFW clusters in a survey area of $50$~deg$^2$ similar to the XXL survey \citep{XXL}. The stacked tangential shear profile $\langle g_+\rangle(R)$ and the number counts in redshift bins $\Delta N(z)$ are used as summary statistics for both methods. By performing a series of forward simulations, we have obtained convergent solutions for the posterior distribution from both methods. We find that \textsc{abc-pmc} recovers broader posteriors than \textsc{pydelfi}, especially for the $\Omega_\mathrm{m}$ parameter. \textsc{pydelfi} recovers convergent posteriors from an order of magnitude fewer simulations than \textsc{abc-pmc}. For a weak-lensing survey with a source density of $n_\mathrm{g}=20$~arcmin$^{-2}$, we find posterior constraints on $S_8=\sigma_8(\Omega_\mathrm{m}/0.3)^{0.3}$ of $0.836\pm 0.032$ and $0.810\pm 0.019$ from \textsc{abc-pmc} and \textsc{pydelfi}, respectively.

Throughout this study, we have made several simplifying assumptions in our forward simulations, particularly in using a single NFW halo description of cluster lenses (see Section~\ref{subsec:assumptions}). In our forthcoming work, we  will improve our simulator by implementing more realistic models of galaxy clusters and weak-lensing noise properties.

The analysis framework developed in this study will be particularly powerful for cosmological inference with ongoing cluster cosmology programs, such as the \textit{XMM}-XXL survey \citep{XXL} and the \textit{eROSITA} all-sky survey \citep{Brunner2021}, in combination with wide-field weak-lensing surveys. Simulation tools developed in this study will also be implemented into the publicly available \textsc{skypy} package \citep{Amara2021,2020zndo...3755531S}.

\acknowledgments
We thank the anonymous referee for constructive suggestions and comments. We thank I-Non Chiu and Ka-Hou Leong for providing helpful comments and suggestions. We also thank Richard P. Rollins and Lucia F. de la Bella for their valuable help to implement part of the code developed in this work into the \textsc{skypy} package. We are grateful to the SkyPy collaboration for fruitful discussions. This work is supported by the Ministry of Science and Technology of Taiwan (grants MOST~106-2628-M-001-003-MY3 and MOST~109-2112-M-001-018-MY3) and by the Academia Sinica Investigator award (grant AS-IA-107-M01).\\
%%
%\software{abcpmc \citep{abcpmc}, Astropy \citep{2018AJ....156..123A}, Colossus \citep{2018ApJS..239...35D}, emcee \citep{emcee}, matplotlib \citep{Matplotlib}, NumPy \citep{numpy}, pathos \citep{2012arXiv1202.1056M, pythos}, PyDelfi \citep{2019MNRAS.488.4440A}, pygtc \citep{pygtc}, Python \citep{python3}, Scipy \citep{scipy}, TensorFlow \citep{tensorflow2015-whitepaper} }

%%% arXiv@@
\textit{Software}:
abcpmc \citep{abcpmc}, Astropy \citep{2018AJ....156..123A}, Colossus \citep{2018ApJS..239...35D}, emcee \citep{emcee}, matplotlib \citep{Matplotlib}, NumPy \citep{numpy}, pathos \citep{2012arXiv1202.1056M, pythos}, PyDelfi \citep{2019MNRAS.488.4440A}, pygtc \citep{pygtc}, Python \citep{python3}, Scipy \citep{scipy}, TensorFlow \citep{tensorflow2015-whitepaper}

\bibliography{tam2021}{}

%%%%%%%%%%%%%%%%%%%%%%%%%%%%%%%%%%%%%%%%%%%%%%%%%%%%%%%%%%%%%%%%%%%
%%%%%%%%%%%%%%%%%%%%%%%%%%%%%%%%%%%%%%%%%%%%%%%%%%%%%%%%%%%%%%%%%%%
%%%
%%% Appendix
%%%
%%%%%%%%%%%%%%%%%%%%%%%%%%%%%%%%%%%%%%%%%%%%%%%%%%%%%%%%%%%%%%%%%%%
%%%%%%%%%%%%%%%%%%%%%%%%%%%%%%%%%%%%%%%%%%%%%%%%%%%%%%%%%%%%%%%%%%%

\clearpage

\begin{appendix}

\section{Mass Scaling Relations}
\label{sec:mscale}

In this appendix, we outline a general procedure for modeling the mass scaling relations in cosmological inference. As evident from Equation~(\ref{eq:dndz}), modeling of the cluster abundance requires the knowledge of the survey selection function and the observable--mass relation. In practical applications, the selection function can be determined from observations and calibrated by simulations. External observations can be used to constrain the observable--mass relation, which sets the ``mass scale'' of the cluster sample detected by a survey.
 
Weak lensing offers a robust but scattered proxy $M_\mathrm{500,WL}$ for the true halo mass $M_{500}$ of galaxy clusters. Forward modeling allows for a statistical mass calibration by linking the cluster observable $O_{500}$ (e.g., X-ray luminosity) and the weak-lensing mass $M_\mathrm{500,WL}$ through the following coupled, scattered equations \citep[e.g.,][]{Umetsu2020hsc}:
\begin{equation}
\label{eq:scaling}
 \begin{aligned}
  \log{O_{500}} &= \alpha + \beta\log{M_{500}} + \gamma\log{E(z)} \pm \sigma_\mathrm{int}(\log O),\\
  \log{M_{\mathrm{500,WL}}} &= \log(1+b_\mathrm{WL}) + \log{M_{500}}\pm \sigma_\mathrm{int}(\log M_\mathrm{WL}),
 \end{aligned}
\end{equation}
where $E(z)\equiv H(z)/H_0$ is the dimensionless Hubble function, $\alpha, \beta$, and $\gamma$ are the intercept, mass-trend, and redshift-trend parameters for the $O$--$M$ relation; $\sigma_\mathrm{int}(\log O)$ and  $\sigma_\mathrm{int}(\log M_\mathrm{WL})$ are the lognormal intrinsic dispersions at fixed $M_{500}$ for $O_{500}$ and $M_\mathrm{500,WL}$, respectively; and $b_\mathrm{WL}$ characterizes the bias in weak-lensing mass estimates. In the first line of Equation~(\ref{eq:scaling}), all normalization constants are absorbed into the intercept $\alpha$. The intrinsic scatter in $M_\mathrm{WL}$ of $\sim 20\percent$, or $\sigma_\mathrm{int}(\log M_\mathrm{WL})\sim 0.2/\ln{10}$, arises from fluctuations of the cluster lensing signal due to intrinsic variations associated with halo assembly bias and asphericity. In a general procedure, some of the parameters describing the mass scaling relations (Equation~(\ref{eq:scaling})) should be let free in forward modeling.

When a scattered $M_\mathrm{WL}$--$M$ relation (i.e., $\sigma_\mathrm{int}(\log{M_\mathrm{WL}})> 0$) is used to generate weak-lensing masses $M_\mathrm{WL}$ for individual NFW clusters, the synthetic weak-lensing signal $\langle g_+^\mathrm{sim}\rangle(R|\bp)$ (Equation~(\ref{eq:gtmod})) should be created using their weak-lensing masses $M_\mathrm{500,WL}$ instead of $M_{500}$. In this way, we effectively account for the effect of intrinsic scatter due to halo assembly bias and asphericity, without creating triaxial halos with scattered concentrations (see Section~\ref{subsec:assumptions}).

In this study, we have largely simplified our modeling procedure to facilitate the interpretation of results. Specifically, we take the cluster observable $O_{500}$ as a hypothetical unbiased mass proxy $M^\prime_{500}$ (i.e., $\alpha=0, \beta=1, \gamma=0$); for weak lensing, we assume that weak lensing provides an unbiased mass calibration with zero astrophysical scatter (i.e., $b_\mathrm{WL}=0, \sigma_\mathrm{int}(\log M_\mathrm{WL})=0$):
\begin{equation}
 \label{eq:obs2mass}
 \begin{aligned}
  \log M^\prime_{500} &= \log M_{500}\pm \sigma_\mathrm{int},\\
  M_\mathrm{500,WL} &= M_{500},\\
 \end{aligned}
\end{equation}
with $\sigma_\mathrm{int}=0.1/\ln{10}$. 

Because of the simplified mass-scaling relations, we use in this work the normalization $A$ of the minimum mass function $M_\mathrm{lim}(z)$ (Equation~(\ref{eq:mmin})) to represent this mass-scale degree of freedom for the survey sample.

\section{Tests with Toy Model Simulations}
\label{sec:toymodel}

In this appendix, we consider two simplified toy models to demonstrate the utility of likelihood-free approaches based on forward simulations. Here we neglect the scatter between the true and observable cluster mass (i.e., $M'=M$) and fix the number of selected clusters $N_\mathrm{cl}$ (i.e., no statistical fluctuation and no Eddington bias). These toy models thus reduce to a mass calibration problem. In addition to the forward-modeling methods described in Section~\ref{sec:methods}, we also employ a conventional maximum-likelihood (ML) approach based on a single-mass-bin NFW fit to the stacked lensing signal \citep[e.g.,][]{Umetsu2020rev}. 
\\

\paragraph*{\rm\textbf{Toy model~\RN{1}}}:
For the first toy model, we assume a Dirac delta mass function $\delta_\mathrm{D}(M_{200}-M_{200}^*)$ with $M_{200}^*=10^{14}h^{-1}M_\odot$ at a single cluster redshift of $z=0.3$. Here $M^*$ is the only parameter of this model that sets the cluster mass scale. We create a synthetic weak-lensing dataset for a sample of $N_\mathrm{cl}=150$ clusters using the forward-modeling procedure described in Section~\ref{sec:methods}. For parameter inference, we use an uninformative uniform prior of $\log(M_{200}^*/h^{-1}M_\odot)\in[12,16]$. Since we consider only Gaussian shape noise as a source of statistical fluctuations in this analysis, the resulting uncertainty on the single parameter $M_{200}$ is expected to scale as $1/\sqrt{n_\mathrm{g}}$ regardless of the inference methods. To examine this scaling of the errors, we will consider an additional noisy realization of synthetic weak-lensing observations with $n_\mathrm{g}=1$~galaxies~arcmin$^{-2}$.

\paragraph*{\rm{\textbf{Toy model~\RN{2}}}}:
For the second toy model, we assume that the cluster mass $M_{200}$ is lognormally distributed with a mean logarithmic mass of $\mu= \langle\log(M_{200}/h^{-1}M_\odot) \rangle$ and a logarithmic dispersion of $\sigma_{\log M_{200}}$. We model the redshift distribution of clusters with a generalized gamma distribution of the form: %\citep[e.g.,][]{Seitz1996}:  
\begin{equation}
\label{eq:gammafunction}
\begin{aligned}
\frac{dN_\mathrm{cl}}{dz} &=
    \frac{\beta N_\mathrm{cl}}{\Gamma\left[(1+\alpha)/\beta\right]}
    \left(\frac{z}{z_1}\right)^{\alpha}\exp\left[-\left(\frac{z}{z_1}\right)^{\beta}\right]\frac{1}{z_1}, \\
%    A&=\frac{1}{\Gamma\left[(1+\alpha)/\beta\right]},\\
    z_1&=z_0\frac{\Gamma\left[(1+\alpha)/\beta\right]}{\Gamma\left[(2+\alpha)/\beta\right]},
\end{aligned}
\end{equation}
where $z_0$ is the mean cluster redshift and $N_\mathrm{cl}$ is the total number of clusters. In this model, we set $\alpha=2$, $\beta=4$,  $z_0 = 0.3$, and $N_\mathrm{cl}=150$. We assume that the mean logarithmic mass $\mu$ of the mass probability distribution function $P(\log{M_{200}})$ evolves with redshift as \citep{Sereno2016,Umetsu2020hsc}\footnote{The mass probability distribution $P(\log{M})$ of observable-selected clusters is mainly shaped by the following two effects: first, as predicted by the cosmic mass function, more massive objects are less abundant; second, less massive objects tend to be fainter and more difficult to detect. Accordingly, $P(\log{M})$ tends to be unimodal, and it evolves with redshift \citep{Sereno2016}.
}
\begin{align} 
    \mu(z)=\mu_0+\gamma_0\log{\left[\frac{D_\mathrm{L}(z)}{D_\mathrm{L}(z_0)}\right]},
    \label{eq:Mdistribution}
\end{align}
where $D_\mathrm{L}(z)$ is the luminosity distance at redshift $z$, $\mu_0$ is the mean at the reference redshift $z=z_0$, and $\gamma_0$ describes the redshift trend of $\mu(z)$. In this toy model, we have three parameters in total, namely, $\mu_0$, $\sigma_{\log M_{200}}$, and $\gamma_0$. In this model, we set $\mu_0=14$, $\sigma_{\log M_{200}}=0.7/\ln{10}\approx 0.30$, and $\gamma_0=0.5$.

To construct a sample of clusters, we draw a set of 150 redshifts and masses from the respective distributions (Equations~(\ref{eq:gammafunction}) and (\ref{eq:Mdistribution})) and produce a synthetic weak-lensing dataset using our simulator. We use uninformative uniform priors for the three parameters: $\mu_0\in[12, 16]$, $\sigma_{\log M_{200}}\in[0.5/\ln{10},1.0/\ln{10}]$, and $\gamma_0\in[0.2,0.9]$. In the conventional method, we only consider the (effective) mass scale of the sample as a single parameter, which is extracted from the stacked lensing signal $\langle g_+\rangle(R)$.\\

Here we briefly summarize our inference procedures for three different approaches. 

\begin{enumerate}
\item In the \textsc{abc-pmc} analysis, parameter sets sampled from the prior distribution are compared to the stacked lensing profile $\langle g_+\rangle(R)$ from synthetic weak-lensing data according to Equation~(\ref{eq:d1}). To obtain convergent results, a series of iterations are preformed. In each iteration, $10^3$ accepted samples are generated.  We use the stopping criterion defined in Section~\ref{sec: abc} (Figure~\ref{fig:abc_flowchart}). The total number of simulations required for convergence is larger than $\mathcal{O}(10^6)$. 

\item  For the \textsc{pydelfi} analysis, the neural network architecture is an ensemble of six neural density estimators, including one masked autoregressive flow with five masked autoencoders for density estimations,  each with 2 hidden layers of 50 hidden units, and five mixture density networks with $1,2,\dots,5$ Gaussian components respectively, each with 2 hidden layers of 30 hidden units. We use nonlinear activation functions, $\tanh$, for all neural units. We divide the inference task into 20 training steps with 1000 simulations for an initial training step.  A batch-size of 800 in each training cycle is also given. Ten percent of simulations are set as a validation sample to avoid over-fitting. 

\item In the conventional approach, we fit the stacked cluster lensing signal $\langle g_+\rangle(R)$ with a single NFW profile. Here the logarithm of the mass scale $\log{(M_{200}/h^{-1}M_\odot)}$ of the sample is a single parameter of interest. The concentration parameter is set according to the mean $c$--$M$ relation $c_{200}(M_{200},z)$ of \citet{Diemer19}. We derive the posterior probability distribution of the effective mass scale using the \textsc{emcee} python package \citep{emcee}. For the mass-scale parameter, we use the same uniform prior as in the forward-modeling cases, $\log(M_{200}/h^{-1}M_\odot)\in [12, 16]$.
\end{enumerate}

The Gaussian likelihood for the conventional approach is given by
\begin{equation}
\label{eq:likelihood}
\begin{aligned}
    -2\ln\mathcal{L} = \sum_{i=1}^{N_\mathrm{bin}}\left\{
    \frac{\left[
    \langle g_{+}\rangle(R_i) - \widehat{g}_{+}(R_i|\bp)\right]^2}{\sigma_{g,i}^2} 
    +\ln\left(2\pi\sigma_{g,i}^2\right)\right\},
\end{aligned}
\end{equation}
where $i$ runs over all radial bins, $\langle g_+\rangle(R_i)$ is the stacked reduced tangential shear in the $i$th bin, $\sigma_{g,i}$ is its measurement uncertainty, and $\widehat{g}_+(R_i)$ is the expectation value predicted by the model $\bp$.

%%% Table 3
\begin{deluxetable}{lccc}
\tablecolumns{4}
\tablewidth{0pt}
\tabletypesize{\scriptsize}
\tablecaption{\label{tab:toymodel1}
Posterior Summaries for Toy Model~\RN{1}}
\tablehead{
\colhead{Survey sensitivity} & 
\multicolumn{3}{c}{$M_{200}^*$ ($10^{14}h^{-1}M_\odot$)}\\ \cline{2-4} 
\colhead{} & 
\multicolumn{1}{c}{Maximum-likelihood } & 
\multicolumn{1}{c}{ABC-PMC} & 
\multicolumn{1}{c}{PYDELFI} 
}
\startdata
$n_\mathrm{g}=1$~arcmin$^{-2}$  &     $1.07\pm 0.19$& $0.99\pm0.25$ & $1.09\pm 0.18$ \\
$n_\mathrm{g}=20$~arcmin$^{-2}$  &    $0.98\pm0.04$ & $0.98\pm0.05$ & $0.98 \pm 0.04$ \\
$n_\mathrm{g}=400$~arcmin$^{-2}$ &    $1.00\pm0.01$ & $1.00\pm0.01$  & $1.00 \pm 0.01$ 
\enddata
\end{deluxetable}

%%% Table 4 
\begin{deluxetable}{lccccccc}
\tablecolumns{8}
\tablewidth{0pt}
\tabletypesize{\scriptsize}
\tablecaption{\label{tab:toymodel2}
 Posterior Summaries for Toy model~\RN{2}
}
\tablehead{
 \colhead{Survey sensitivity} & 
 \multicolumn{1}{c}{Maximum-likelihood} & 
 \multicolumn{3}{c}{ABC-PMC}  &  
 \multicolumn{3}{c}{PYDELFI}\\   \cline{2-8} 
 \colhead{} &
 \multicolumn{1}{c}{$\log(M_\mathrm{200}/h^{-1}M_\odot)$} &
 \multicolumn{1}{c}{$\mu_0$} &
 \multicolumn{1}{c}{$\sigma_{\log M_\mathrm{200}}$ } &
 \multicolumn{1}{c}{$\gamma_0$} &
 \multicolumn{1}{c}{$\mu_0$} &
 \multicolumn{1}{c}{$\sigma_{\log M_\mathrm{200}}$ } &
 \multicolumn{1}{c}{$\gamma_0$} 
}
\startdata
$n_\mathrm{g}=20$~arcmin$^{-2}$  & $13.98 \pm 0.02$ & $13.95 \pm 0.05$ & $0.77\pm 0.15$ & $0.57\pm 0.21$ & $13.97 \pm 0.05$ & $0.73 \pm 0.10$ & $0.56 \pm 0.18$\\
$n_\mathrm{g}=400$~arcmin$^{-2}$ & $14.00 \pm 0.00$\tablenotemark{a} & $13.99 \pm 0.04$ & $0.71\pm 0.13$ & $0.55\pm 0.20$ & $13.99 \pm 0.02$ & $0.74 \pm 0.09$ & $0.49 \pm 0.12$ 
\enddata
\tablecomments{The values of the input parameters are $\mu_0:=\langle\log(M_{200}(z=z_0)/h^{-1}M_\odot) \rangle=14$, $\sigma_{\log M_{200}}=0.7/\ln{10}\approx 0.30$, and $\gamma_0=0.5$.}
\tablenotetext{a}{The $1\sigma$ uncertainty is less than $0.01$.}
\end{deluxetable}

Table~\ref{tab:toymodel1} summarizes the resulting posterior constraints on the mass-scale parameter $M_{200}^*$ of Toy model~\RN{1} obtained using the three different methods. In each survey depth ($n_\mathrm{g}$), we find that the resulting posterior constraints on $M^*_{200}$ using the \textsc{abc-pmc}, \textsc{pydelfi}, and ML methods are consistent with each other, except for the noisiest realization with $n_\mathrm{g}=1$~galaxies~arcmin$^{-2}$ where \textsc{abc-pmc} recovers a broader posterior distribution than from the other two methods. This is expected because only in the limit of $\epsilon\rightarrow0$, the accepted samples are drawn from the exact posterior. For a noisier realization that requires an ABC rejection sampling over a wider parameter space, it will take longer for the iterative process to converge.
Otherwise, the magnitude of errors approximately scales as $1/\sqrt{n_\mathrm{g}}$, as expected. We note that for each survey depth ($n_\mathrm{g}$), we analyze only one particular realization of synthetic observations created using our forward simulator. As a result, the posterior means can be deviated from the ground truth. However, all methods are expected to show a consistent shift in the posterior mean.

Table~\ref{tab:toymodel2} shows the results for Toy model~\RN{2}. For the conventional ML method, we obtain a constraint on the parameter $\langle\log{M_{200}}\rangle$ similar to that for Toy model~\RN{1}.  This is because in this approach we only consider the effective mass scale extracted from the stacked lensing signal. The ML method, however, yields a much narrower posterior distribution compared to the marginal errors in $\mu_0$ obtained with \textsc{abc-pmc} and \textsc{pydelfi}. 

Overall, \textsc{abc-pmc} and \textsc{pydelfi} yield comparable constraints on the model parameters. We notice again that \textsc{abc-pmc} produces somewhat broader posterior distributions compared to \textsc{pydelfi}. 
%Such a broadening of posterior distributions is generally expected for \textsc{abc-pmc}, because only in the limit of $\epsilon\rightarrow0$ the ABC algorithm will draw samples from the true posterior distribution. 
For any nonzero $\epsilon$, an ABC approximate posterior is always broader than the true posterior distribution. Compared to \textsc{abc-pmc} that requires $\mathcal{O}(10^6)$ forward simulations, \textsc{pydelfi} is computationally much more efficient as it requires only $\mathcal{O}(10^5)$ forward simulations.

It is worthy to point out that for \textsc{abc-pmc} and \textsc{pydelfi}, the errors in $\mu_0$ do not simply scale with the number density of background galaxies. This is because, in addition to shape noise, forward simulations properly account for the statistical fluctuations of the cluster sample drawn from the distributions given by Equations~(\ref{eq:gammafunction}) and (\ref{eq:Mdistribution}). In particular, for the survey depth of $n_\mathrm{g}=400$~galaxies~arcmin$^{-2}$, the uncertainty in the mass-scale parameter is dominated by this sample variance, which is neglected in the ML method. In contrast, all relevant sources of statistical fluctuations are automatically taken into account in \textsc{abc-pmc} and \textsc{pydelfi} based on forward simulations.

From these results, we find that for a simple problem (Toy model~\RN{1}) in which the underlying likelihood is well described by a Gaussian, both likelihood-free and ML approaches obtain an unbiased recovery of the model parameters. For a more complex problem (Toy model~\RN{2}), forward-modeling approaches can properly account for all relevant statistical effects, which are properly encoded in the resulting posterior distributions, and thus improve completeness and accuracy of the analysis. We note that forward modeling assuming a Gaussian likelihood \citep[e.g.,][]{Sereno2016,Chiu2021} is also capable of properly handling such statistical effects, as long as the Gaussian assumption is valid. We also notice that compared to the ABC approach, \textsc{pydelfi} requires much less simulations and produces narrower posteriors \citep[see][]{2018MNRAS.477.2874A,2018PhRvD..98f3511L}.

\end{appendix}

\end{document}